\documentclass[12pt]{article}
\input{maths}
\newcommand{\blind}{1}
\usepackage{xr}
\begin{document}

\def\spacingset#1{\renewcommand{\baselinestretch}%
{#1}\small\normalsize} \spacingset{1}

%%%%%%%%%%%%%%%%%%%%%%%%%%%%%%%%%%%%%%%%%%%%%%%%%%%%%%%%%%%%%%%%%%%%%%%%%%%%%%

\if1\blind
{
  \title{\bf Partially-shared Imaging Regression 
  on Integrating Heterogeneous Brain-Cognition Associations across Alzheimer’s Diagnoses}
  \author{Yang Sui$^{1}$, Qi Xu$^{2}$, Ting Li$^{3}$, Yang Bai$^{3}$, and Annie Qu$^{4}$\thanks{Corresponding author: \href{mailto:aqu2@ucsb.edu}{\textcolor{black}{aqu2@ucsb.edu}}.} 
 \hspace{.2cm} \\
	\textit{$^{1}$Department of Statistics and Data Science,}\\
\textit{University of California, Los Angeles}\\
\textit{$^{2}$Department of Statistics and Data Science, Carnegie Mellon University}\\ 
	\textit{$^{3}$School of Statistics and Data Science,}\\
\textit{Shanghai University of Finance and Economics}\\
	\textit{$^{4}$Department of Statistics and Applied Probability,}\\
    \textit{University of California, Santa Barbara}} 
  \date{}
  \maketitle
} \fi

\if0\blind
{
  \bigskip
  \bigskip
  \bigskip
  \begin{center}
  {\LARGE\bf Partially-shared Imaging Regression %with Smooth Spatial Component Integration for 
  on Integrating Heterogeneous Brain-Cognition Associations across Alzheimer’s Diagnoses}
\end{center}
  \medskip
} \fi

\bigskip
\begin{abstract}
Alzheimer’s Disease Neuroimaging Initiative (ADNI) diagnostic groups present strong heterogeneous associations among demographic, imaging, and cognitive data. We propose a novel PArtially-shared Imaging Regression (PAIR) model to represent imaging coefficients as weighted combinations of smooth spatial components. A Total Variation penalty is applied to enforce spatial smoothness, and a Selective Integration penalty is introduced to adaptively learn partial-sharing structures across groups. Theoretically, we establish minimax-optimal error bounds that dynamically adapt to varying sharing paradigms. Numerically, PAIR achieves predictive accuracy comparable to advanced deep learning models while providing superior interpretability. Applied to ADNI data, PAIR reveals substantial heterogeneity in brain-cognition pathways between cognitively normal (CN) and cognitively impaired (CI) groups, with hippocampal imaging contributing minimally in the CN group but substantially in the CI group, particularly in the CA1, CA3, and presubiculum subfields.
\end{abstract}

\noindent%
\vfill

\newpage
\spacingset{1.9} % DON'T change the spacing!
\section{Introduction}
Alzheimer’s disease (AD) is a chronic neurodegenerative condition characterized by progressive cognitive decline and behavioral impairments due to the degeneration of brain cells \citep{jack2010hypothetical}. Identifying risk factors that influence AD progression is critical for enabling early intervention and targeted treatment strategies \citep{li2022regression,li2024partially}. 
However, existing research often faces challenges from limited sample sizes within specific data sources, potentially leading to imprecise estimation and prediction \citep{ghosh2025ensemble}. While demographic and imaging factors influencing cognitive performances may share common patterns across different data sources \citep{shan2024merging}, considerable heterogeneity is expected due to differences in disease stage and study population \citep{xu2025representation,sui2026multi}. 
For instance, cognitively normal (CN) and cognitively impaired (CI) individuals may exhibit distinctly different associations between demographic/imaging covariates and cognitive outcomes \citep{bourzac2025women,peng2015correlation}.

The primary goal of this paper is to integrate imaging and demographic information from multiple data sources within the Alzheimer’s Disease Neuroimaging Initiative (ADNI) study to map the biological pathways associated with AD-related cognitive outcomes \citep{yu2022mapping,li2024partially}. To characterize these pathways, we utilize baseline demographic covariates and hippocampal surface imaging data, focusing on CN and CI baseline diagnostic groups. By investigating how these covariates influence cognitive performances across distinct baseline statuses, we can better identify group-specific risk factors related to cognitive decline and inform targeted interventions \citep{he2023new}.

\begin{figure}[t]
    \centering
    \includegraphics[width=0.8\linewidth]{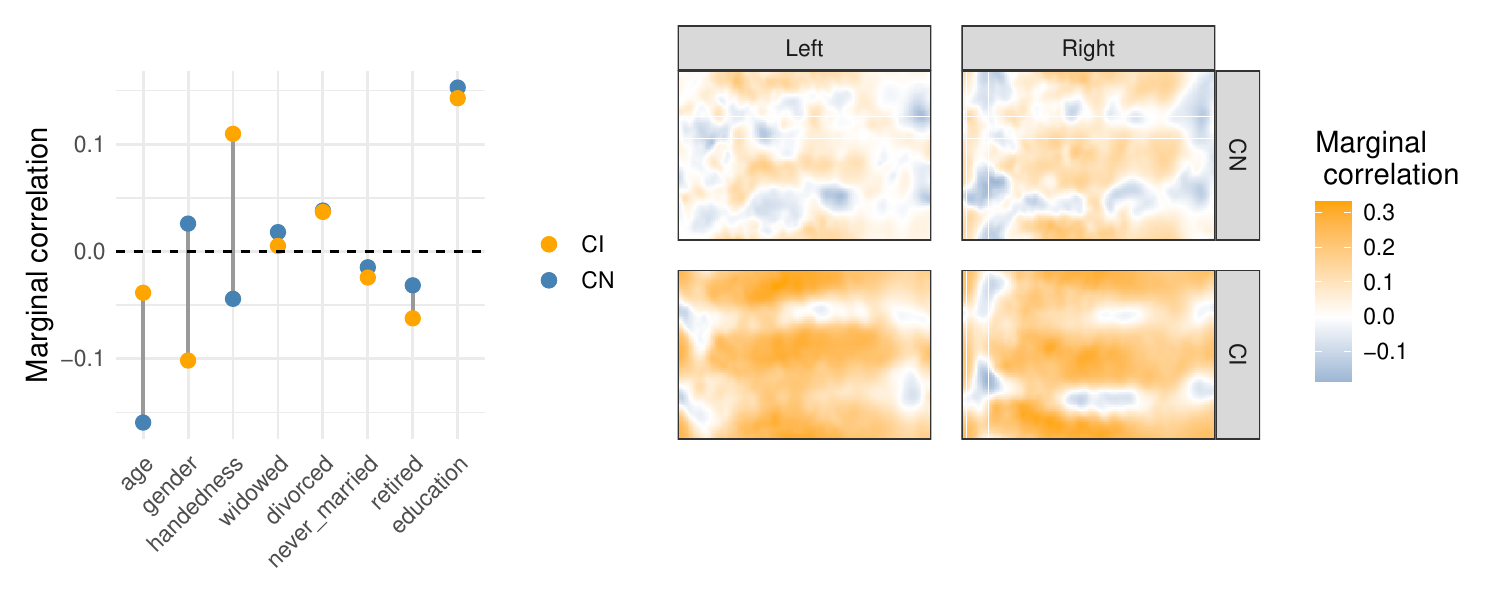}
    \vskip -0.2in
    \caption{The left panel shows the marginal correlations between each demographic covariate and the response across diagnostic groups (CN and CI), while the right panel presents the marginal correlations between each pixel and the response across brain hemispheres (Left and Right) and diagnostic groups (CN and CI).}
    \label{fig:marginal correlations}
\end{figure}

To motivate our approach, we explicitly explore the marginal correlations between the covariates (including hippocampal imaging pixels) and cognitive scores, stratified by diagnostic group. As illustrated in Figure~\ref{fig:marginal correlations}, while some correlation patterns are shared between the CN and CI groups, distinct differences and group-specific heterogeneity are highly evident. This complex, partially-shared structure poses new analytical challenges. First, traditional statistical methods based on linear decompositions are inadequate for learning complex imaging coefficients. Second, existing integration methods rarely accommodate the partially-shared spatial structures inherent in imaging data. Thus, there is a pressing need in developing novel statistical methodologies capable of effectively mapping these heterogeneous pathways and achieving imaging data integration.

To address these challenges, we propose a PArtially-shared Imaging Regression (PAIR) model with smooth spatial component integration to learn heterogeneous imaging coefficients across different data sources. In particular, we assume that each imaging coefficient can be expressed as a weighted combination of a set of smooth spatial components. By jointly learning this spatial component set and the corresponding weights, PAIR adaptively identifies spatial components which contribute to group-specific imaging coefficient, thereby capturing the partially-shared structure across sources.
To more effectively estimate the smooth spatial components, we apply a Total Variation (TV) penalty to each component \citep{wang2017generalized}. We also apply a Selective Integration Penalty (SIP) to the weight matrix motivated by \citep{xu2025representation}, which encourages adaptive learning of partially-shared components across imaging coefficients.

Previous studies have employed methods that either directly vectorize imaging data into a single vector \citep{zhou2014brain} or adopt functional representations of images \citep{li2022regression, li2024partially}. However, both approaches inevitably disrupt the intrinsic spatial structure of the image \citep{feng2021brain}.
Our proposed PAIR method is closely related to the growing literature on imaging regression models and their variants, motivated by the increasing availability of high-dimensional imaging data \citep{zhou2013tensor, wang2017generalized, li2018tucker, wu2023sparse,bai2024generalized}. While these methods preserve the spatial structure of imaging data, they do not explicitly account for the piecewise smoothness of imaging coefficients. Alternatively, \citet{wang2017generalized} proposed a two-dimensional TV regularization to encourage spatial smoothness. While recent data integration methods like the group penalty framework by \citet{ma2024multi} can jointly analyze correlated images, they disrupt spatial structures and fail to accommodate partially-shared patterns across sources.
In contrast, our proposed PAIR framework flexibly captures a broad spectrum of sharing structures, including single-task learning, fully shared, and partially-shared settings.

Theoretically, we establish a non-asymptotic upper bound for the average imaging coefficient estimation error of the proposed PAIR estimator. Notably, this error bound explicitly quantifies the statistical complexity across various sharing structures. The effective sample size contributes to the error bound adaptively: it scales with the total pooled data in a fully shared regime, reduces to the individual sample size in single-task learning, and interpolates between these two extremes under partial sharing. Furthermore, we derive a matching minimax lower bound, proving that our proposed estimator achieves the optimal estimation rate.

Numerically, we validate the effectiveness of our proposed PAIR against various statistical baseline methods and advanced deep learning models across synthetic simulations, the ChestMNIST imaging dataset \citep{yang2023medmnist}, and the ADNI real data application. The results demonstrate that PAIR consistently outperforms existing statistical methods in both prediction and estimation, particularly when the degree of sharing among imaging coefficients is high. Furthermore, even when compared to complex deep learning approaches, our PAIR achieves comparable prediction performance while providing highly interpretable insights into the underlying data structures and mechanisms.

Finally, our application of PAIR to the ADNI datasets reveals substantial heterogeneity in the brain-cognition associations between the CN and CI groups. While demographic covariates contribute to cognitive scores in both groups, the additional predictive contribution of hippocampal imaging is minimal in the CN group but several times higher in the CI group. For the CN group, the estimated effects of demographic covariates remain largely consistent after adjusting for hippocampal imaging; in contrast, notable changes are observed for the CI group, suggesting that covariate effects on cognition are modulated by hippocampal structural changes. The strongest imaging signals in the CI group are concentrated in the CA1, CA3, and presubiculum subfields, consistent with established patterns of AD-related hippocampal damage. Detailed findings are presented in Section~\ref{sec: real data}.

\section{Data and Problem Description for ADNI}\label{sec: data}
The ADNI is a large-scale study designed to collect clinical, imaging, and cognitive data to evaluate interventions and track the progression of AD \citep{mueller2005ways}. We construct two datasets from the ADNI database (\href{https://adni.loni.usc.edu}{adni.loni.usc.edu}), encompassing cognitively normal subjects (CN) and cognitively impaired subjects (CI). 
The baseline scalar covariates (summarized in Table~\ref{tab:adni_summary}) include age, gender, handedness, three marital-status indicators (widowed, divorced, and never married), retirement status, and years of education. For each subject, we use the image processing pipeline in \citet{li2007hippocampal} to compute hippocampal morphometry surface measures, resulting in a $100 \times 150$ matrix, where each element represents the radial distance from a specific coordinate on the hippocampus surface to the medial core. Since the left and right hippocampi have two-dimensional radial distance measures and may exhibit asymmetry \citep{li2024partially}, we analyze them separately.

To measure cognitive impairment, we utilize the mini-mental state examination (MMSE) score \citep{folstein1975mini}. Here, the scalar covariates and hippocampal surface data are collected at baseline, and the MMSE score is measured 12 months later, focusing our analysis on how baseline factors relate to future cognitive status. Summary statistics for the two datasets are reported in Table~\ref{tab:adni_summary}. As expected, the CI group has a substantially lower 12-month MMSE score than the CN group, reflecting more severe cognitive impairment.

\begin{table}[t]
\renewcommand{\arraystretch}{0.6}
\centering
\caption{Summary statistics for the ADNI CN and CI datasets.}
\vskip -0.2in
\label{tab:adni_summary}
\resizebox{0.6\textwidth}{!}{
\begin{tabular}{lcc}
\hline
Characteristic & CN & CI \\
\hline
Sample size & 182 & 424 \\
Age & 76.104 (5.152) & 75.425 (7.168) \\
Education & 16.225 (2.680) & 15.500 (2.970) \\
Male / Female & 97 / 85 & 264 / 160 \\
Right-handed / Left-handed & 169 / 13 & 393 / 31 \\
Widowed & 28 & 47 \\
Divorced & 13 & 19 \\
Never married & 11 & 7 \\
Married & 130 & 351 \\
Retired / Not retired & 150 / 32 & 347 / 77 \\
12-month MMSE & 29.242 (1.039) & 25.264 (3.701) \\
\hline
\end{tabular}
}
\end{table}

Although there has been extensive research on the effects of demographic covariates and hippocampus imaging data on cognitive MMSE scores, few studies have considered integrating various diagnosis datasets. Given the complex structures identified in our initial exploration (Figure~\ref{fig:marginal correlations}), we are interested in the following scientific questions:
\begin{itemize}
    \item (Q1) Given limited sample sizes within each group, can shared brain-cognition associations be leveraged to enhance future cognitive prediction?
    \item (Q2) How do baseline scalar and imaging covariates contribute to future cognitive decline across diagnostic groups, and how can we quantify and identify shared and group-specific effect patterns?
\end{itemize}

We propose the PArtially-shared Imaging Regression (PAIR) model to address (Q1)–(Q2) for heterogeneous imaging data sources from CN and CI groups.

\section{Methodology}\label{sec: method}
%\subsection{Preliminaries}
We consider $T$ heterogeneous imaging data sources (each corresponding to a diagnostic group). For the $t$-th data source, we model the response as
\begin{align*}
    Y_t = \langle Z_t, \beta_t \rangle + \langle X_t, C_t \rangle + \varepsilon_t,
\end{align*}
where $\{(Y_t, Z_t, X_t)\}_{t \in [T]}$ denote the responses, covariates, and images collected from $T$ sources. The notation $\langle \cdot, \cdot \rangle$ refers to the inner product between two vectors of the same dimension or two matrices of the same size, and $\{1,\ldots,T\}$ is denoted as $[T]$ for brevity. For each $t$, the vector $Z_t\in\RR^d$ represents the covariate, the matrix $X_{t} \in \RR^{p\times q}$ represents the image of size $p\times q$, and $Y_{t}$ signifies the scalar response. The parameters $\beta_t \in \mathbb{R}^d$ and $C_t \in \mathbb{R}^{p \times q}$ are group-specific coefficients associated with the covariates and images, respectively, and the noise term $\varepsilon_t$ is assumed to be independent of both $Z_t$ and $X_t$.

Given $T$ heterogeneous imaging data sources, the sets of covariate and imaging coefficients, $\{\beta_t\}_{t\in[T]}$ and $\{C_t\}_{t\in[T]}$, may exhibit both shared patterns and specific heterogeneity. For example, as shown in Figure~\ref{fig:marginal correlations}, covariates and hippocampus surfaces have partially-shared effects on cognitive scores across CN and CI groups in the ADNI study. In this paper, we assume that the covariate coefficients $\{\beta_t\}_{t\in[T]}$ are fully heterogeneous  since their estimation is relatively straightforward, and their integration can be readily handled by existing integration approaches developed for linear models \citep{li2022transfer,duan2023adaptive}. Instead, our focus is on the integration of partially-shared imaging coefficients $\{C_t\}_{t\in[T]}$, a setting for which there is limited work addressing heterogeneity with partial-sharing in image-based models.

\subsection{PArtially-shared Imaging Regression (PAIR)}% with Smooth Spatial Component Integration}
In particular, the coefficient matrices $C_t$ across different  sources could share similar structures, which enable us to integrate $T$ sources to improve the estimation of $C_t$'s, leading to enhanced estimation and signal detection across all sources. We propose the following decomposition on $C_t$'s:
\begin{align}\label{eqn: partial shared}
    C_t = \sum_{r=1}^Rw_{tr}B_r,
\end{align}
where $C_t$ is a combination of $R$ spatial components $\{B_r \in \RR^{p\times q}\}_{r \in[R]}$ and $W = (w_{tr})_{T\times R} \in \RR^{T\times R}$ collects weights across all sources for all spatial components. In the following exposition, $W_{t\cdot} = (w_{t1}, \cdots, w_{tR})$ are used to denote the weight vector for the $t$-th data source. The decomposition (\ref{eqn: partial shared}) adopts shared spatial components such that all sources can be used to estimate $\{B_r\}_{r\in[R]}$ if the corresponding weights are non-zero, which increase the effective sample size for estimating $C_t$'s. On the other hand, the weight matrix $W$ indicates the sharing structures across sources. There are several sharing structures that are considered in the literature can be accommodated by $W$: (1) single-task learning (STL): if $\text{supp}(W_{t_1\cdot})\cap \text{supp}(W_{t_2\cdot}) = \emptyset$ for any $t_1, t_2\in[T]$, indicating there is no sharing spatial component across any two datasets. In this scenario, (\ref{eqn: partial shared}) reduces to STL and there is no benefit of data integration; (2) fully-shared (FS) structure: if $\text{supp}(W_{t\cdot}) = [R]$ for all $t\in[T]$, implying all sources employ all spatial components. Even though the magnitude of weights can be different, data integration can be effective by estimating $\{B_r\}_{r \in[R]}$ with all samples from $T$ sources; (3) partially-shared (PS) structure \citep{gaynanova2019structural}: If there exists some $t_1, t_2\in[T]$ and $\text{supp}(W_{t_1\cdot}) \cap \text{supp}(W_{t_2\cdot}) \neq \emptyset$, illustrating some spatial components are shared by some sources, which is the most general sharing structure among aforementioned ones. Since $W$ is also unknown and requires to be estimated, the sharing structure assumptions such as (1)-(2) on $W$ could be misspecified. Therefore, we consider the partially-shared structure in this work.

\begin{figure}[t]
    \centering
\includegraphics[width=0.65\linewidth]{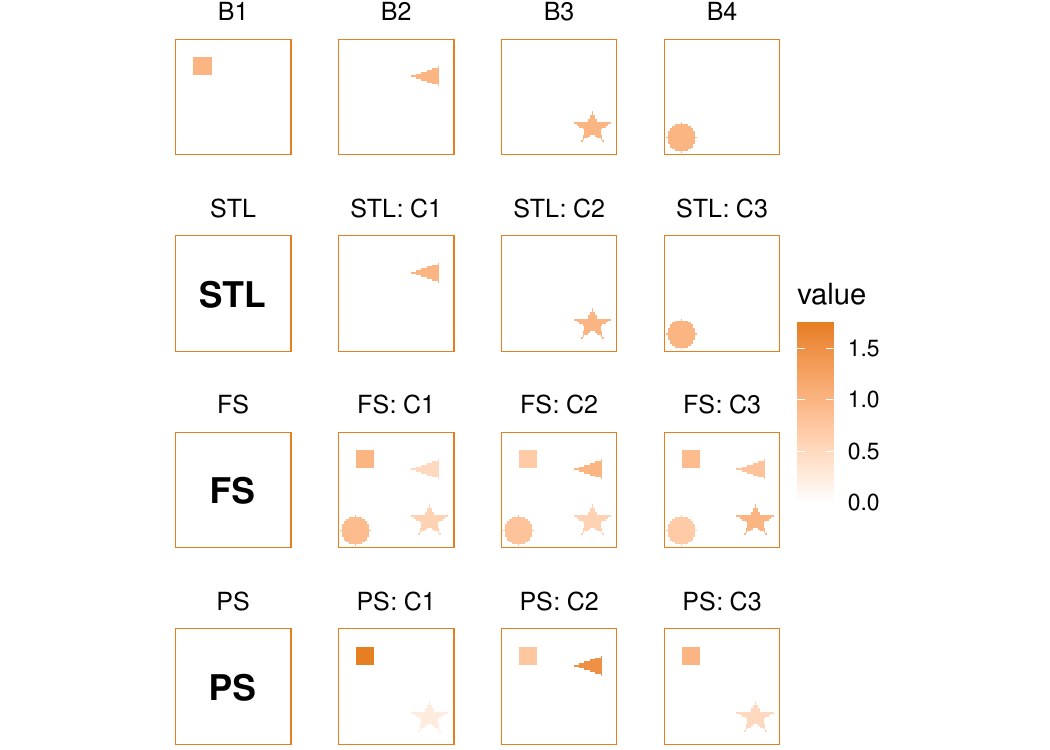}
    \vskip -0.2in
    \caption{Spatial component set and imaging coefficients under different sharing settings.}
    \label{fig:true coefficients}
\end{figure}

We provide a toy example in the following to illustrate the differences among the aforementioned sharing structures. We consider $T = 3$ sources and $R = 4$ spatial components, each corresponding to a distinct spatial pattern: square, triangle, star, and circle, as shown in the first row of Figure~\ref{fig:true coefficients}. Four different weight matrix configurations are examined. In the STL setting, each imaging coefficient contains only one unique smooth spatial component without any shared components across sources (second row of Figure~\ref{fig:true coefficients}). In the FS setting, the weight matrix $W$ is dense, and all imaging coefficients incorporate all spatial components, indicating a fully shared representation with varying weights (third row of Figure~\ref{fig:true coefficients}).  In the PS setting, the weight matrix $W$ is sparse, resulting in limited sharing: the square component is shared by all sources, the star component is shared by $C_1$ and $C_3$, and the triangle component is unique to $C_2$ (fourth row row of Figure~\ref{fig:true coefficients}). Our proposed method is designed to flexibly capture these varying degrees of shared structure among imaging coefficients across different sources.
% W1tensor([[0.7181, 0.7856, 1.4419],
%        [1.0151, 1.1001, 1.4262],
%        [1.0525, 0.7051, 1.2449]])

% W2 tensor([[1.2500, 1.0000, 2.0000],
%        [0.0000, 0.2500, 0.5000],
%        [1.5000, 0.7500, 1.7500]])

% W3 tensor([[1.7500, 0.0000, 0.2500],
%        [0.7500, 1.5000, 0.0000],
%        [1.0000, 0.0000, 0.5000]])
To accommodate complex patterns in $B_r$'s in ~\eqref{eqn: partial shared} and flexible partially-shared structures among $W_{t\cdot}$'s, we consider additional techniques to control the complexity of $B_r$'s and the degree of partially-shared structure in $W_{t\cdot}$' explicitly. To address the first issue, we consider applying the following Total Variation (TV) penalty \citep{wang2017generalized, wu2024individualized} to all spatial components $\{B_r\}_{r\in[R]}$:
\begin{align}
\label{eq:tv-demo}
\|B_r\|_{\tv}:=
\sum_{1\le j_1\le p,\;1\le j_2\le q}
\Big(
|B_{r,j_1,j_2}-B_{r,j_1,j_2-1}|
+
|B_{r,j_1,j_2}-B_{r,j_1-1,j_2}|
\Big),
\end{align}
where $B_{r,j_1,0}=B_{r,j_1,1}$ and $B_{r,0,j_2}=B_{r,1,j_2}$. 
The TV form~\eqref{eq:tv-demo} is referred to the $\ell_1$-norm, or anisotropic version of TV. An $\ell_2$-norm, or isotropic version of TV is defined similarly. This TV penalty encourages the spatial components to remain smooth in space, effectively capturing non-linear features of the spatial components while maintaining sparsity and local structure. This approach helps to address the challenge of modeling complex, non-linear spatial components accurately. %Compared to directly applying the TV penalty to the complete set of imaging coefficients $ \{C_t\}_{t\in[T]} $, applying the TV penalty to the signal candidate set $\{B_r\}_{r\in[R]}$ has clear computational advantages. This is because the signal candidates are inherently simpler than the full imaging coefficients, making the optimization process more straightforward and efficient.

To address the second issue, we consider applying the following Selective Integration Penalty (SIP) motivated by \citep{xu2025representation} on the weight matrix $W$ to enforce the integration of the smooth spatial components $\{B_r\}_{r\in[R]}$:
\begin{align}\label{eqn: P(W)}
Q_r(W):=\sum_{t=1}^T \mathds{1}({|w_{tr}|\not=0}),\quad        P(W):=
\sum_{r=1}^R \min\!\Big(1,\frac{T-Q_r(W)}{T-1}\Big),
\end{align}
where $\mathds{1}(\cdot)$ is the indicator function. In~\eqref{eqn: P(W)},
$Q_r(W)$ counts how many sources actually share $B_r$.
If $B_r$ is shared by all sources ($Q_r(W)=T$), then the penalty is $0$ and $B_r$ is not discouraged.
If $B_r$ appears in at most one source ($Q_r(W)\le 1$), then the penalty saturates at $1$.
For partial sharing levels ($2\le Q_r(W)\le T-1$), the penalty decreases as $Q_r(W)$ increases, so optimization is encouraged to activate $B_r$ in more sources rather than keeping it source-isolated.
Summing over $r\in[R]$, $P(W)$ selectively discourages columns that are supported on only a few sources and thereby promotes integration of the shared components $\{B_r\}_{r\in[R]}$ across sources. However, $P(W)$ in~\eqref{eqn: P(W)} involves an indicator function $\mathds{1}(\cdot)$, which makes directly optimizing the term computationally challenging. To address this issue, we propose the following smoothed SIP on the weight matrix $ W $:
\begin{equation}
\label{eq:Ptau-demo}
Q_r(\tau;W):=\sum_{t=1}^T \min\!\Big(\frac{|w_{tr}|}{\tau},1\Big),
\quad
P_\tau(W)
\;:=\;
\sum_{r=1}^R \min\!\Big(1,\frac{T-Q_r(\tau;W)}{T-1}\Big).
\end{equation}
In~\eqref{eq:Ptau-demo}, the parameter $ \tau $ controls the threshold of the weight $ w_{tr} $. Originally, we push $ w_{tr} \neq 0 $ to indicate that the smooth spatial component $ B_r $ is shared by $ C_t $, but now we modify it such that only when $ |w_{tr}| $ exceeds the threshold $ \tau $, the smooth spatial component $ B_r $ is considered to be shared by $ C_t $, which aligns better with practical scenarios. Based on~\eqref{eq:Ptau-demo}, $P_\tau(W)$ in~\eqref{eq:Ptau-demo} can be viewed as a smoothed version of the original penalty~\eqref{eqn: P(W)}, making it computationally easier to optimize.

\subsection{Estimation and Implementation}
Suppose that we observe $T$ data sources $\{\{Y_{ti},Z_{ti},X_{ti}\}_{i=[n_t]}\}_{t\in[T]}$, the squared loss for the $t$-th data source can be written as $\ell_t(\beta_t,C_t)=(2n_t)^{-1}\sum_{i=1}^{n_t}(Y_{ti}-\langle Z_{ti},\beta_t\rangle-\langle X_{ti},C_t\rangle)^2$.
To handle the classification problem, we can replace the square loss with cross entropy loss without further modification. We then propose the following TV and smoothed SIP penalized minimization problem:
\begin{align}\label{eqn: minimize}
\begin{split}
        \{\{\widehat\beta_t\}_{t\in[T]},\{\widehat B_r\}_{r\in[R]},\widehat W\}=\argmin_{\{\beta_t\}_{t\in[T]},\{B_r\}_{r\in[R]},W} \frac{1}{T}\sum_{T}\ell_t\left(\beta_t,\sum_{r=1}^Rw_{tr}B_r\right)\\+\lambda\sum_{r=1}^R\|B_r\|_{\tv}+\gamma P_\tau(W),
\end{split}
\end{align}
where $\frac{1}{T}\sum_{T}\ell_t(\sum_{r=1}^Rw_{tr}B_r)$ is the average squared loss across all $T$ sources, $\lambda$ and $\gamma$ are the parameters corresponding to the TV penalty~\eqref{eq:tv-demo} and smoothed SIP~\eqref{eq:Ptau-demo}, and $\tau$ is the parameter in~\eqref{eq:Ptau-demo}. The higher the value of $ \lambda $, the more it constrains the complex structure of the learned spatial components. Similarly, the higher the values of $ \gamma $ and $ \tau $, the more they constrain the degree of signal integration. The optimization problem in~\eqref{eqn: minimize} can be solved alternatively between
covariate coefficient set $\{\beta_t\}_{t\in[T]}$ and imaging coefficient set $\{\{B_r\}_{r\in[R]},W\}$: Given a fixed set $\{\beta_t\}_{t\in[T]}$, $\{\{B_r\}_{r\in[R]},W\}$ can be updated through a
gradient-based algorithm; given a fixed set $\{\{B_r\}_{r\in[R]},W\}$, minimizing~\eqref{eqn: minimize} is equivalent to solving linear regression problem
for each source. In~\eqref{eqn: minimize}, fusion penalties such as $\sum_{s,t} \|\beta_s - \beta_t\|$ can be incorporated to encourage shared patterns among the covariate coefficients $\{\beta_t\}_{t\in[T]}$, or alternatively, similarity-based penalties as in \citet{duan2023adaptive}.

In practice, the initial values for all parameters can be chosen from independent random Gaussian matrices with mean 0 and small variance. We use the automatic differentiation implemented in PyTorch \citep{paszke2019pytorch} along with the Adam optimizer \citep{kingma2014adam} to update imaging coefficient set $\{\{B_r\}_{r\in[R]},W\}$. For simplicity, we use the same learning rate in all steps. 
For hyperparameter tuning, the proposed method includes multiple hyperparameters: optimization-related hyperparameters (e.g., penalty parameters $\lambda,\gamma,\tau$, learning rate, patience of early stopping). We choose the grid search method \citep{yu2020hyper} and select the best hyperparameter setting if it attains the smallest loss on an independent validation set. 

The computational complexity of~\eqref{eqn: minimize} is analyzed as follows.
Let $N=\sum_{t=1}^T n_t$ be the total sample size.
Computing the linear covariate part $\langle Z_{ti},\beta_t\rangle$
for all $(t,i)$ has cost $O(Nd)$.
Computing the imaging part $\langle X_{ti},B_r\rangle$ for all
$(t,i,r)$ has cost $O(NRpq)$.
Evaluating the TV penalty over $\{B_r\}_{r=1}^R$ costs $O(Rpq)$.
The smoothed SIP penalty $P_\tau(W)$ in~\eqref{eq:Ptau-demo} costs $O(TR)$.
Consequently, the overall cost of one forward--backward pass for updating
$\{B_r\}_{r\in[R]}$ and $W$ is $
O(Nd + NRpq + Rpq + TR)
= O(N(d+Rpq))$,
since $N$ and $pq$ are typically large while $R$ and $T$ are moderate.
For a fixed set of tuning parameters $(\lambda,\gamma,\tau)$, a first-order method
with $K$ iterations therefore has overall complexity $O(KN(d+Rpq))$.

\section{Theoretical Guarantees}
\label{sec:pair-theory}
This section presents theoretical guarantees for the proposed PAIR imaging estimator, with covariates $Z_{ti}$ and linear effects $\beta_t$ omitted for clarity. We derive an upper bound for the average coefficient estimation error under the constrained PAIR framework, which holds over a broad range of sharing structures. Furthermore, we establish a matching minimax lower bound over an induced coefficient class, demonstrating that our upper bound is minimax-rate optimal up to logarithmic factors and the image dimensions.

For simplicity, we assume that each source has $n$ samples. For each $(t,i)\in[T]\times[n]$, $Y_{ti} =\langle X_{ti}, C_t^\ast\rangle+ \varepsilon_{ti}$, where
$C_t^\ast\in\RR^{p\times q}$ is the unknown imaging coefficient. We assume that the coefficients admit a shared-component representation $
C_t^\ast=\sum_{r=1}^R w_{tr}^\ast B_r^\ast$ for $t\in[T]$, where $B_r^\ast\in\RR^{p\times q}$ are shared spatial components, and
$w_{tr}^\ast\in\RR$ are source-specific weights.
Write $B^\ast=(B_1^\ast,\dots,B_R^\ast)$ and $W^\ast=(w_{tr}^\ast)_{t,r}$.

For any candidate $(B,W)$ with $B=(B_1,\dots,B_R)$ and $W=(w_{tr})_{t,r}$, define $\ell(B,W):=
\frac{1}{2nT}\sum_{t=1}^T\sum_{i=1}^n
(Y_{ti}-\langle X_{ti}, \sum_{r=1}^R w_{tr}B_r\rangle)^2$.
Define the $R\times R$ Gram matrices $
G_W(W):=\frac1T W^\top W$ and $G_B(B):= (\langle B_r,B_s\rangle_F)_{r,s\in[R]}$. Fix constants $m_W>0$, $m_B>0$, $M_W>0$, $M_B>0$,
Define the classes $\mathcal W:=\{W\in\RR^{T\times R}: \|W\|_\infty\le M_W, \lambda_{\min}(G_W(W))\ge m_W\}$ and $\mathcal B:=\{B=(B_1,\dots,B_R): \max_{r}\|B_r\|_F\le M_B, \lambda_{\min}(G_B(B))\ge m_B\}$, where $\lambda_{\min}(G)$ denotes the smallest
eigenvalue for a symmetric matrix $G$.
Given tuning parameters $\lambda>0$ and $\gamma>0$, we define
\begin{equation}
\label{eq:pair-est-constrained-demo}
(\widehat B,\widehat W)
\in
\argmin_{(B,W)\in\mathcal B\times\mathcal W}
\Big\{
\ell(B,W)
+
\lambda\sum_{r=1}^R \|B_r\|_{\tv}
+
\gamma\,P_\tau(W)
\Big\},\quad \widehat C_t:= \sum_{r=1}^R \widehat w_{tr}\widehat B_r,
\end{equation}
for $t\in[T]$. 
The constrained classes $\mathcal B$ and $\mathcal W$ are imposed to make the bilinear map
$(B,W)\mapsto {C_t=\sum_{r=1}^R w_{tr}B_r}_{t\in[T]}$ well-conditioned. Without restrictions,
the factorization is non-identifiable: $(B,W)$ and $(BA,WA^{-1})$ yield the same coefficients
${C_t}$ for any invertible $A\in\RR^{R\times R}$, so $C_t$ can be held fixed while the factors
become arbitrarily ill-scaled, for instance, $\|B_r\|_F\to\infty$ and $\|W\|_\infty\to 0$.
The constraints $\|W\|_\infty\le M_W$ and $\max_r\|B_r\|_F\le M_B$ exclude such scaling degeneracies.
Moreover, $\lambda_{\min}(G_B(B))\ge m_B>0$ prevents near-linear dependence among
$\{B_r\}_{r=1}^R$, while $\lambda_{\min}(G_W(W))\ge m_W>0$ ensures the columns of $W$ are neither
nearly collinear nor nearly zero across sources. We also assume that the true $(B^\ast,W^\ast)\in\mathcal{B}\times\mathcal{W}$. For each $r\in[R]$, let $S_r$ denote the set of adjacent pairs in $B_r^\ast$ with non-zero differences. Let $s_r:=|S_r|$ and define $s_B:=\sum_{r=1}^R s_r$.
In the following, we proceed with some essential assumptions:
\begin{assumption}
\label{ass:noise-design-demo}
Conditional on $\{X_{ti}\}$, the noises $\{\varepsilon_{ti}\}$ are independent,
mean-zero, and sub-Gaussian with variance proxy at most $\sigma^2$: $\EE[\exp(u\varepsilon_{ti})\mid X_{ti}]
\le
\exp(\frac12\sigma^2u^2)$ for all $u\in\RR$.
Moreover, $\|X_{ti}\|_F \le M_X$, almost surely, for all $(t,i)$.
\end{assumption}

\begin{assumption}
\label{ass:re-demo}
There exists $\kappa>0$ such that for all collections $C_1,\dots,C_T\in\RR^{p\times q}$,
$\frac1{nT}\sum_{t=1}^T\sum_{i=1}^n \langle X_{ti},C_t\rangle^2\ge\kappa \cdot \frac1T\sum_{t=1}^T \|C_t\|_F^2$.
\end{assumption}
Assumption~\ref{ass:noise-design-demo} imposes standard regularity conditions on the noise terms and imaging matrix $X_{ti}$. Assumption~\ref{ass:re-demo} specifies the restricted eigenvalue condition, which is commonly employed in regularized estimation. See~\citep{negahban2012unified} for more details.

\begin{theorem}
\label{thm:pair-main-demo}
Let $c, C_1, C_2, C_3 > 0$ be universal constants independent of $(n, p, q, R, T, \sigma)$. Under Assumptions~\ref{ass:noise-design-demo} and \ref{ass:re-demo}, let $\{\widehat C_t\}_{t\in[T]}$ be any global minimizer of \eqref{eq:pair-est-constrained-demo}. For any fixed $\delta \in (0,1)$, suppose the tuning parameters satisfy
\begin{equation}
\label{eq:tuning-demo}
\lambda \ge 8c\, M_W M_X \sigma \, pq \sqrt{\frac{\log(2Rpq/\delta)}{nT}},
\qquad
\frac{\gamma}{\tau} \ge \frac{8c(T-1)M_B M_X \sigma}{T}\sqrt{\frac{\log(2TR/\delta)}{n}}.
\end{equation}
Then with probability at least $1-\delta$, the average coefficient estimation error satisfies
\begin{equation}
\label{eq:rate-demo}
\frac1T\sum_{t=1}^T\|\widehat C_t-C_t^\ast\|_F^2
\le
\frac{C_1 s_B \lambda^2}{\kappa^2 m_W}
+
\frac{C_2 R T^2 \gamma^2}{\kappa^2 m_B (T-1)^2 \tau^2}.
\end{equation}
In particular, plugging the parameter scalings from \eqref{eq:tuning-demo} into \eqref{eq:rate-demo} yields
\[
\frac1T\sum_{t=1}^T\|\widehat C_t-C_t^\ast\|_F^2
\le
\frac{C_3}{\kappa^2} \left( \frac{s_B M_W^2 M_X^2 \sigma^2 p^2 q^2}{m_W} \frac{\log(2Rpq/\delta)}{nT} + \frac{R M_B^2 M_X^2 \sigma^2}{m_B} \frac{\log(2TR/\delta)}{n} \right).
\]
\end{theorem}
Choose $m_W=\lambda_{\min}(G_W(W^\ast))$, where $m_W$ quantifies the amount of collective weight energy available across sources, meaning how well the source weights jointly excite the shared $B$-components. Accordingly, up to logarithmic factors the $B$-block term in \eqref{eq:rate-demo} behaves like $\frac{s_B}{nT}\cdot\frac{1}{m_W}$, with effective sample size $nT\,m_W$. In a partially-shared structure where column $r$ is supported on a set $\cS_r\subset[T]$ of size $|\cS_r|$, one typically has the scaling $m_W\asymp \min_r |\cS_r|/T$ because $(G_W(W^\ast))_{rr}=T^{-1}\sum_t(w_{tr}^\ast)^2\asymp |\cS_r|/T$ and the minimum eigenvalue is governed by the least-excited column; substituting yields $\frac{s_B}{nT}\cdot\frac{1}{m_W}\asymp \frac{s_B}{n\min_r |\cS_r|}$.

The effective sample size adapts to the sharing regime. In the FS regime, $|\cS_r|\asymp T$ for all $r$, hence $m_W\asymp 1$ and the $B$-block term achieves the pooled rate $\asymp s_B/(nT)$. In the STL regime, $|\cS_r|\asymp 1$, hence $m_W\asymp 1/T$ and the term reverts to the single-task scale $\asymp s_B/n$. In the PS regime where $|\cS_r|\asymp k$ with $1\ll k\ll T$, one has $m_W\asymp k/T$ and the term becomes $\asymp s_B/(nk)$, interpolating between $s_B/n$ and $s_B/(nT)$.
The second term in \eqref{eq:rate-demo} provides the $W$-block contribution, which up to logarithmic factors behaves like $R/n$ and represents the statistical cost of learning source-specific weights. The PAIR can also be extended to integrate the weights $W_{t\cdot}$ across sources, which may further improve the rate when weight-sharing structures are present.

We next establish a minimax lower bound for estimating the coefficient matrices
$\{C_t^\ast\}_{t=1}^T$ under the PAIR model.
Define the induced coefficient class
$\mathcal C
:=\{\{C_t^\ast\}_{t=1}^T:\exists (B^\ast,W^\ast)\in\mathcal B\times\mathcal W,\ C_t^\ast=\sum_{r=1}^R w_{tr}^\ast B_r^\ast,\ \forall t,\ |\cup_{r=1}^R S_r|\le s_B\}$
as the collection of coefficient sequences $C^\ast=\{C_t^\ast\}_{t=1}^T$. Then we have the following results.

\begin{theorem}
\label{thm:minimax-lb-pair}
Assume $R \le T$ and let $M_W$ be a sufficiently large fixed constant. We further assume $R+\lfloor s_B/8\rfloor \le \lceil pq/2\rceil$, where $\lfloor \cdot \rfloor$ and $\lceil \cdot \rceil$ denote the floor and ceiling functions, respectively. This condition ensures that the $p\times q$ spatial grid can accommodate at least $R+\lfloor s_B/8\rfloor$ mutually non-adjacent locations
Then there exists a covariate distribution $\mathbb P_X$ supported on
$\{X:\|X\|_F\le M_X\}$ and an absolute constant $c>0$ such that
\[
\inf_{\widetilde C}\sup_{C^\ast\in\mathcal C}
\mathbb E\left[\frac1T\sum_{t=1}^T \|\widetilde C_t-C_t^\ast\|_F^2\right]
\ge
\max\left(\frac{c\sigma^2s_B}{nT m_W},\frac{c\sigma^2R}{n}\right),
\]
where the infimum ranges over all estimators
$\widetilde C=\{\widetilde C_t\}_{t=1}^T$ measurable with respect to
$\{(X_{ti},Y_{ti})\}_{t\in[T],i\in[n]}$.
\end{theorem}
The upper bound in Theorem~\ref{thm:pair-main-demo} establishes an average estimation error of order $\sigma^2\big(\frac{s_B}{nT m_W} + \frac{R}{n}\big)$, where the two terms represent the distinct statistical costs of recovering the shared spatial components and estimating source-specific weights, respectively. Correspondingly, Theorem~\ref{thm:minimax-lb-pair} provides a matching minimax lower bound of order $\max\big(\frac{\sigma^2s_B}{nT m_W}, \frac{\sigma^2R}{n}\big)$, which quantifies the inherent difficulty of the induced coefficient class $\mathcal{C}$. Given that these two bounds coincide up to logarithmic factors and $(p,q)$, our analysis demonstrates that the proposed PAIR estimator is minimax-rate optimal. Furthermore, the parameter $m_W$ characterizes the transition in statistical complexity as the model interpolates between the fully-shared, partially-shared, and single-task regimes.

\section{Simulation}\label{sec: simu}
In this section, we carry out simulation studies to examine the empirical performance of the proposed method. Specifically, we compare the proposed PAIR model with several competitors: voxelwise regression (VR), which conducts linear regression of the response on each voxel separately for each data source, and its regularized form with a Lasso penalty (RVR); tensor regression (TR); regularized tensor regression with a Lasso penalty imposed on the factor vectors (RTR), as proposed in \citet{zhou2013tensor}, applied separately to each data source; scalar-on-image regression via a TV penalty (SIRTV), as proposed in \citet{wang2017generalized}, applied separately to each data source; and pooled scalar-on-image regression with a TV penalty (POOL), which assumes that all data sources share the same imaging coefficient. In addition, we include three deep learning baselines: CNN, a convolutional neural network baseline for jointly modeling image and covariate information \citep{lecun1998gradient}; HPS, a hard-parameter-sharing architecture that learns shared image features across sources and then combines them with covariates for prediction \citep{ruder2017overview}; and PRE, a pretrained baseline based on ResNet18, where pretrained image features are extracted and then combined with covariates through a prediction head \citep{he2016deep}.

We consider the following data generation process (DGP),
$Y_{ti} = \langle Z_{ti}, \beta_t \rangle + \langle X_{ti}, C_t \rangle + \varepsilon_{ti}$,
where $Z_{ti}$ is a $d$-dimensional covariate vector with entries independently sampled from the standard normal distribution, and $X_{ti}$ is a $p \times q$ matrix generated according to the simulation setup described below, with $d = 5$ and $p = q = 64$. The elements of the coefficient vectors $\{\beta_t\}_{t \in [T]}$ and the random errors $\varepsilon_{ti}$ are also independently drawn from the standard normal distribution. We consider $T=3$ imaging data sources and assume that $\{C_t\}_{t\in[3]}$ follows the shared structure $C_t=\sum_{r=1}^Rw_{tr}B_r$ for $t\in[3]$, where we choose $R=3$ smooth spatial components $\{B_r\}_{r\in[3]}$. We evaluate three settings for the weight matrix $W=(w_{tr})$. In Setting 1, its entries are drawn from $U[0.5, 1.5]$. In Setting 2, the entries are randomly permuted from a sequence of evenly spaced values between $0$ and $2$. Setting 3 follows the same generation mechanism as Setting 2, except that one randomly chosen entry in each row is exactly set to $0$.
Setting 1 corresponds to a scenario in which the smooth spatial components are more widely shared across different data sources. In Setting 2, the degree of sharing is smaller. In Setting 3, the sharing is even more limited: no single source-specific coefficient $C_t$ contains all smooth spatial components, and each source includes at most $R-1$ components. We visualize the spatial components and examples of the true coefficients under the three settings in Supplementary Material~\ref{supp: simu data}. 

We generate $n_1$, $n_2$, and $n_3$ samples $\{\{Y_{ti},Z_{ti},X_{ti}\}_{i\in[n_t]}\}_{t\in[3]}$ for the three datasets. The data are split into 60\% training, 20\% validation, and 20\% testing sets. We estimate the coefficients using the training data and select the parameters $R$, $\lambda$, $\gamma$, and $\tau$ using the validation data by minimizing the loss $\frac{1}{T}\sum_{t=1}^T\ell_t(\beta_t,C_t)$. For the other competing methods, the hyperparameters are also tuned through validation. The implementation details for all methods are summarized in Supplementary Material~\ref{sec:simu-implementation}.

Based on the selected model, we use $\widehat C_t$ and $\widehat Y_{ti}=\langle Z_{ti},\widehat\beta_t\rangle+\langle X_{ti},\widehat C_t\rangle$ to evaluate prediction performance on the testing data. For estimation, we consider two metrics: average estimation error (AEE$_C$), $\frac13\sum_{t=1}^3\|\widehat C_t-C_t\|_F$, and average estimation error (AEE$_\beta$), $\frac13\sum_{t=1}^3\|\widehat\beta_t-\beta_t\|$. For prediction, we calculate the average root mean square error (ARMSE), $\frac13\sum_{t=1}^3\sqrt{\frac{1}{n_t}\sum_{i=1}^{n_t}(\widehat Y_{ti}-Y_{ti})^2}$, on the testing data.

We construct the imaging covariates $X_{ti}$ using two distinct approaches: generating independent standard normal variables or utilizing real ChestMNIST images \citep{yang2023medmnist}. Due to space constraints, the main text focuses on the ChestMNIST application, with sample sizes set to $n_1=n_2=n_3\in\{600,1200\}$. The corresponding estimation and prediction results are summarized in Table~\ref{tab:medmnist_all_metrics_new}. Results for the fully simulated normal data, which show similar patterns with PAIR achieving the best performance across all settings, are deferred to Supplementary Material~\ref{sec:simu_simulated}.
From Table~\ref{tab:medmnist_all_metrics_new}, we observe that, as the sample size increases, almost all methods improve in both estimation accuracy and predictive performance. In the smaller-sample setting with $n_1=n_2=n_3=600$, the competing methods are more strongly affected by limited sample size, whereas PAIR already shows clear advantages. VR and RVR perform moderately, with VR showing particularly poor predictive performance because it fails to capture the true signals. In contrast, RVR imposes sparsity and therefore improves both estimation and prediction. TR and RTR show a similar pattern: RTR improves upon TR, but both methods still perform substantially worse than the best methods. POOL, by ignoring the heterogeneity among the imaging coefficients, consistently underperforms SIRTV in the ChestMNIST image simulation. The three deep learning baselines, CNN, HPS, and PRE, also do not perform competitively here, especially compared with PAIR. Overall, PAIR achieves the best estimation and predictive performance across all settings.

\begin{table}
\renewcommand{\arraystretch}{0.7}
\scriptsize
\setlength{\tabcolsep}{2.5pt}
\centering
\caption{Comparison of AEE$_C$, AEE$_\beta$, and ARMSE across various methods and ChestMNIST datasets under three settings with different sample sizes. The results are summarized by mean and standard deviation over 100 repeats.}
\vskip -0.2in
\label{tab:medmnist_all_metrics_new}
\resizebox{\textwidth}{!}{
\begin{tabular}{lllcccccccccc}
\hline
Setting & $n$ & Metric & VR & RVR & TR & RTR & SIRTV & POOL & PAIR & CNN & HPS & PRE \\
\hline
\multirow{12}{*}{Setting 1}
& \multirow{6}{*}{600}
& \multirow{2}{*}{AEE$_C$}
& 14.69 & 10.22 & 54.94 & 15.39 & 9.78 & 11.00 & \textbf{5.41} & -- & -- & -- \\
&& & (1.88) & (1.20) & (7.29) & (2.23) & (1.20) & (1.25) & \textbf{(0.64)} & -- & -- & -- \\
&& \multirow{2}{*}{AEE$_{\beta}$}
& 2.02 & 0.87 & 0.67 & 0.44 & 0.47 & 2.43 & \textbf{0.15} & -- & -- & -- \\
&& & (0.38) & (0.19) & (0.15) & (0.09) & (0.12) & (0.56) & \textbf{(0.03)} & -- & -- & -- \\
&& \multirow{2}{*}{ARMSE}
& 24.89 & 5.14 & 6.89 & 4.17 & 4.50 & 25.52 & \textbf{1.47} & 10.30 & 18.19 & 73.39 \\
&& & (6.11) & (0.77) & (0.95) & (0.49) & (0.73) & (7.76) & \textbf{(0.11)} & (1.33) & (2.38) & (8.62) \\
\cline{2-13}
& \multirow{6}{*}{1200}
& \multirow{2}{*}{AEE$_C$}
& 14.62 & 9.68 & 50.22 & 16.30 & 9.24 & 10.54 & \textbf{4.75} & -- & -- & -- \\
&& & (1.91) & (1.15) & (7.54) & (2.89) & (1.19) & (1.13) & \textbf{(0.57)} & -- & -- & -- \\
&& \multirow{2}{*}{AEE$_{\beta}$}
& 2.02 & 0.70 & 0.40 & 0.28 & 0.27 & 2.11 & \textbf{0.09} & -- & -- & -- \\
&& & (0.38) & (0.15) & (0.09) & (0.07) & (0.07) & (0.39) & \textbf{(0.02)} & -- & -- & -- \\
&& \multirow{2}{*}{ARMSE}
& 24.17 & 4.15 & 5.52 & 3.84 & 3.59 & 25.16 & \textbf{1.22} & 7.62 & 13.52 & 63.83 \\
&& & (6.13) & (0.58) & (0.69) & (0.44) & (0.59) & (7.68) & \textbf{(0.06)} & (0.86) & (1.71) & (7.07) \\
\hline
\multirow{12}{*}{Setting 2}
& \multirow{6}{*}{600}
& \multirow{2}{*}{AEE$_C$}
& 16.99 & 11.80 & 55.30 & 16.39 & 11.17 & 16.10 & \textbf{6.47} & -- & -- & -- \\
&& & (0.73) & (0.39) & (7.40) & (2.01) & (0.59) & (1.89) & \textbf{(0.88)} & -- & -- & -- \\
&& \multirow{2}{*}{AEE$_{\beta}$}
& 1.96 & 0.98 & 0.72 & 0.45 & 0.58 & 4.15 & \textbf{0.19} & -- & -- & -- \\
&& & (0.42) & (0.21) & (0.15) & (0.10) & (0.14) & (1.32) & \textbf{(0.03)} & -- & -- & -- \\
&& \multirow{2}{*}{ARMSE}
& 32.72 & 6.36 & 7.08 & 4.15 & 5.43 & 57.30 & \textbf{1.79} & 11.61 & 19.48 & 76.76 \\
&& & (4.82) & (0.49) & (0.89) & (0.69) & (0.47) & (13.31) & \textbf{(0.17)} & (1.20) & (1.73) & (6.37) \\
\cline{2-13}
& \multirow{6}{*}{1200}
& \multirow{2}{*}{AEE$_C$}
& 16.91 & 11.22 & 51.34 & 17.16 & 10.77 & 15.08 & \textbf{5.62} & -- & -- & -- \\
&& & (0.74) & (0.38) & (7.28) & (2.75) & (0.53) & (1.48) & \textbf{(0.66)} & -- & -- & -- \\
&& \multirow{2}{*}{AEE$_{\beta}$}
& 1.96 & 0.79 & 0.42 & 0.29 & 0.35 & 3.07 & \textbf{0.10} & -- & -- & -- \\
&& & (0.42) & (0.17) & (0.10) & (0.07) & (0.08) & (0.88) & \textbf{(0.02)} & -- & -- & -- \\
&& \multirow{2}{*}{ARMSE}
& 31.82 & 5.16 & 5.63 & 3.79 & 4.62 & 56.02 & \textbf{1.33} & 8.51 & 14.72 & 67.31 \\
&& & (4.53) & (0.40) & (0.72) & (0.54) & (0.39) & (13.33) & \textbf{(0.08)} & (0.74) & (1.04) & (5.37) \\
\hline
\multirow{12}{*}{Setting 3}
& \multirow{6}{*}{600}
& \multirow{2}{*}{AEE$_C$}
& 12.57 & 8.85 & 43.31 & 13.02 & 8.30 & 13.66 & \textbf{4.84} & -- & -- & -- \\
&& & (2.37) & (1.59) & (7.58) & (3.00) & (1.60) & (2.52) & \textbf{(1.20)} & -- & -- & -- \\
&& \multirow{2}{*}{AEE$_{\beta}$}
& 1.79 & 0.76 & 0.51 & 0.32 & 0.39 & 4.16 & \textbf{0.17} & -- & -- & -- \\
&& & (0.37) & (0.18) & (0.13) & (0.09) & (0.12) & (1.59) & \textbf{(0.03)} & -- & -- & -- \\
&& \multirow{2}{*}{ARMSE}
& 19.78 & 4.51 & 5.15 & 3.03 & 3.85 & 57.76 & \textbf{1.52} & 10.69 & 15.53 & 55.50 \\
&& & (6.83) & (0.99) & (0.99) & (0.72) & (0.85) & (16.23) & \textbf{(0.21)} & (2.08) & (3.31) & (9.83) \\
\cline{2-13}
& \multirow{6}{*}{1200}
& \multirow{2}{*}{AEE$_C$}
& 12.45 & 8.37 & 42.62 & 12.82 & 7.95 & 13.17 & \textbf{4.23} & -- & -- & -- \\
&& & (2.39) & (1.50) & (7.98) & (2.93) & (1.54) & (2.43) & \textbf{(0.95)} & -- & -- & -- \\
&& \multirow{2}{*}{AEE$_{\beta}$}
& 1.76 & 0.59 & 0.31 & 0.21 & 0.26 & 3.28 & \textbf{0.10} & -- & -- & -- \\
&& & (0.37) & (0.15) & (0.07) & (0.06) & (0.07) & (1.06) & \textbf{(0.02)} & -- & -- & -- \\
&& \multirow{2}{*}{ARMSE}
& 18.93 & 3.64 & 4.20 & 2.75 & 3.30 & 56.89 & \textbf{1.25} & 7.66 & 11.50 & 49.50 \\
&& & (6.59) & (0.74) & (0.76) & (0.62) & (0.71) & (16.07) & \textbf{(0.08)} & (1.33) & (2.15) & (9.10) \\
\hline
\end{tabular}
}
\end{table}

\begin{figure}
    \centering
    \includegraphics[width=0.8\linewidth]{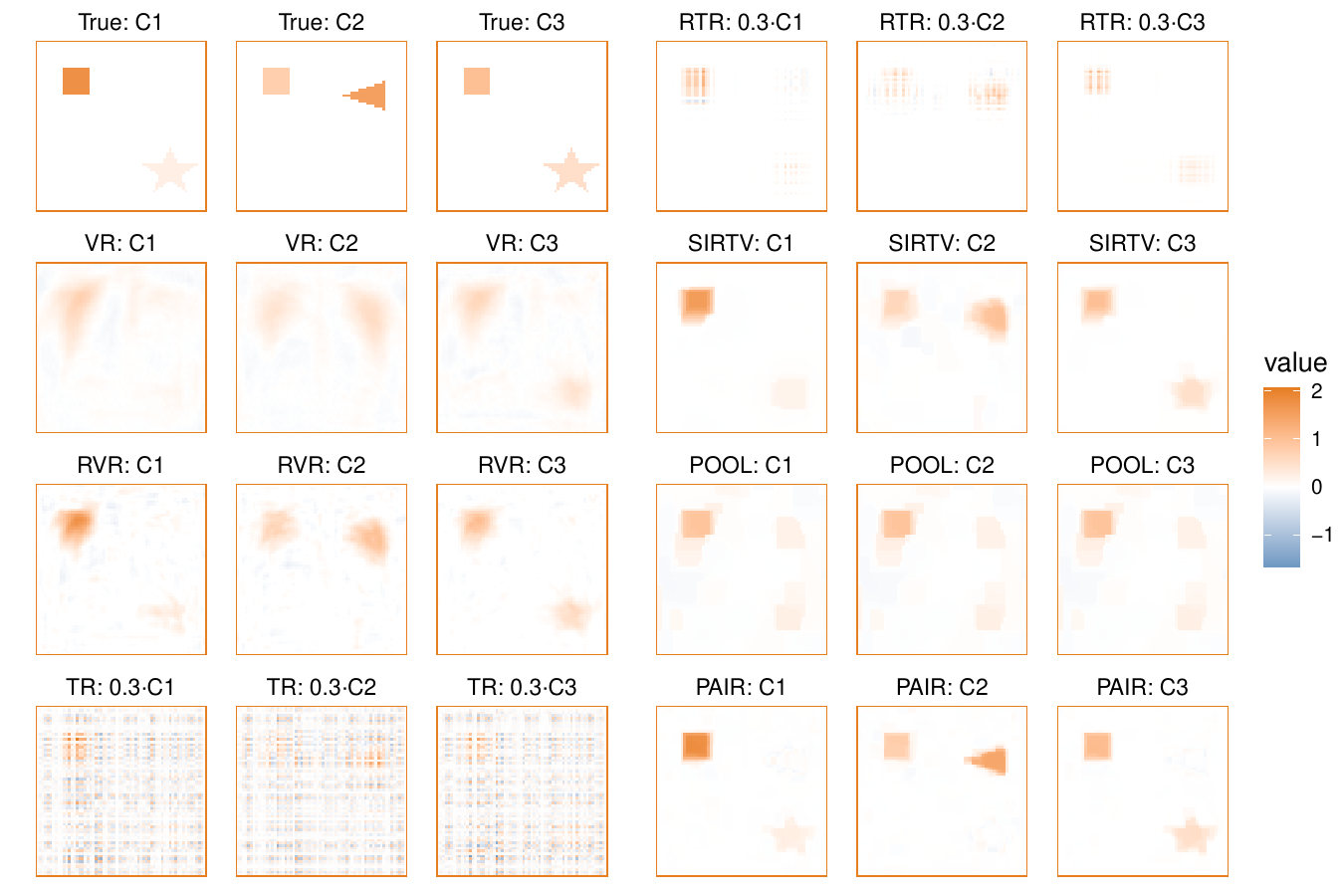}
    \vskip -0.2in
    \caption{True imaging coefficients and the estimated coefficients using different methods and ChestMNIST data with sample size $n_1=n_2=n_3=600$.}
    \label{fig:medmnist_example_plot}
\end{figure}

Furthermore, we visualize the true imaging coefficients and the coefficient estimates obtained by each method in Figure~\ref{fig:medmnist_example_plot}. The methods can be grouped by the quality of their spatial recovery. VR, RVR, TR, and RTR produce noisy estimates that at best roughly capture signal locations: VR and RVR lack spatial structure, while TR and RTR can only recover the linear square signal and fail to identify the triangular and pentagonal patterns. SIRTV and POOL incorporate spatial regularization and capture signal locations more effectively, but SIRTV struggles with precise shapes and edges, and POOL performs worse by ignoring heterogeneity across imaging coefficients. In contrast, our proposed PAIR method accurately estimates both the locations and shapes of the true signals. For example, the edges of the square and triangular signals are captured with high precision. Moreover, PAIR is the only method capable of estimating the complex edges of the pentagonal signal, highlighting the advantages of precise integration.
\section{ADNI Data Analysis}\label{sec: real data}

To address research questions (Q1)--(Q2), we apply the proposed PAIR method to the CN and CI datasets constructed in Section~\ref{sec: data}. The clinical covariates consist of an intercept and demographic variables, specifically age, gender, handedness, three marital-status dummies, retirement status, and years of education. The imaging covariates are the left and right hippocampal surface images, and the response of interest is the MMSE score. Specifically, PAIR is fitted by optimizing the objective in~\eqref{eqn: minimize} for the CN and CI datasets. For comparison, we consider the competing methods introduced in Section~\ref{sec: simu}: VR, RVR, TR, RTR, SIRTV, and POOL. For POOL, we additionally include a disease-stage indicator to account for baseline diagnosis information.

We further evaluate several deep learning and pretrained representation baselines. Specifically, CNN directly learns a predictive model from the hippocampal images and demographic covariates for each dataset separately~\citep{lecun1998gradient}. HPS combines a global image branch and a patch-based summary branch, integrating the resulting image features with demographic covariates for prediction~\citep{ruder2017overview}. PRECAE first learns image representations from all hippocampal images via a convolutional autoencoder, subsequently combining the learned representations with demographic covariates in a downstream ridge regression~\citep{masci2011stacked}. PREMED extracts image representations using  MedImageInsight, a general-domain medical image embedding model~\citep{codella2024medimageinsight}, followed by the identical ridge regression framework. Detailed descriptions of these approaches are provided in Supplementary Material~\ref{sec:adni-dl-pretrained}.

To evaluate predictive performance, we employ repeated five-fold sample splitting. In each repetition, one fold from each data source serves as the testing set. Of the remaining samples, 20\% are allocated as validation data, and the rest for training. For PAIR, the tuning parameters $R$, $\lambda$, $\gamma$, and $\tau$ are selected on the validation set by minimizing the empirical risk $\frac{1}{2}\sum_{t=1}^{2} \ell_t(\beta_t, C_t)$. For all other methods, model selection is similarly based on the validation data according to their respective tuning criteria. Utilizing the selected model, the prediction on the testing data is computed as $\widehat Y_{ti} = \langle Z_{ti}, \widehat\beta_t \rangle + \langle X_{ti}, \widehat C_t \rangle$, and the predictive performance is measured by the testing RMSE, $\sqrt{\frac{1}{n_t}\sum_{i=1}^{n_t}(\widehat Y_{ti}-Y_{ti})^2}$. To make fuller use of the samples, we repeat the five-fold splitting 100 times, resulting in 500 runs for each method under each setting. To assess predictive performance under limited sample sizes, we evaluate the models under both a full-data setting and a 50\%-data setting. For the 50\%-data settings, each repetition independently subsamples 50\% of the subjects from each source prior to the repeated five-fold splitting procedure. The analysis is conducted separately for the left and right hippocampal images. All hyperparameter search spaces and fixed optimization settings are summarized in Supplementary Material~\ref{sec:adni-dl-pretrained}. The robustness of PAIR to the choice of $(R,\lambda,\gamma,\tau)$ within the candidate grid is examined in Supplementary Material~\ref{sec:sensitivity_analysis}, which shows that both predictive performance and the estimated spatial patterns remain stable across hyperparameter selections.
\subsection{Shared Brain-Cognition Associations for Prediction}

The prediction results of PAIR and all competing methods are reported in Table~\ref{tab:adni_rmse}. Several clear patterns emerge from the table.
First, the overall performance of VR, RVR, TR, and RTR is relatively weak across groups and sample-size settings. Moreover, SIRTV consistently outperforms POOL across both diagnostic groups, despite the inclusion of additional diagnostic information in POOL. This suggests that there exists substantial heterogeneity in the relationship between covariates, hippocampus images, and MMSE scores across the CN and CI groups. Ignoring such heterogeneity and directly pooling the data can therefore lead to a clear loss in predictive performance.

\begin{table}[t!]
\renewcommand{\arraystretch}{0.7}
\centering
\caption{Comparison of RMSE across various methods and datasets for left and right hippocampus images under the 50\%-data and full-data settings. The results are based on 500 runs.}
\vskip -0.2in
\label{tab:adni_rmse}
\resizebox{\textwidth}{!}{
\begin{tabular}{cccccccccc}
\hline
~ & ~ & \multicolumn{4}{c}{50\% Data} & \multicolumn{4}{c}{Full Data} \\
\cline{3-10}
~ & Method & \multicolumn{2}{c}{Left} & \multicolumn{2}{c}{Right} & \multicolumn{2}{c}{Left} & \multicolumn{2}{c}{Right} \\
\cline{3-10}
~ & ~ & CN & CI & CN & CI & CN & CI & CN & CI \\
\hline
\multirow{2}{*}{~}
& \multirow{2}{*}{VR}
& 3.709 & 4.598 & 3.715 & 4.667 & 3.725 & 4.635 & 3.760 & 4.664 \\
&
& (0.926) & (0.775) & (0.959) & (0.731) & (0.937) & (0.724) & (0.958) & (0.694) \\
& \multirow{2}{*}{RVR}
& 2.140 & 3.715 & 2.199 & 3.828 & 2.112 & 3.682 & 2.134 & 3.769 \\
&
& (0.404) & (0.625) & (0.470) & (0.578) & (0.291) & (0.434) & (0.313) & (0.394) \\
& \multirow{2}{*}{TR}
& 7.253 & 5.807 & 7.506 & 6.000 & 4.282 & 4.908 & 4.409 & 5.056 \\
&
& (3.908) & (1.705) & (3.857) & (1.845) & (2.464) & (1.189) & (2.372) & (1.274) \\
& \multirow{2}{*}{RTR}
& 2.111 & 4.090 & 2.215 & 4.155 & 1.502 & 3.743 & 1.521 & 3.771 \\
&
& (1.361) & (0.725) & (1.569) & (0.772) & (0.673) & (0.491) & (0.629) & (0.459) \\
& \multirow{2}{*}{SIRTV}
& 1.366 & 3.757 & 1.372 & 3.817 & 1.188 & 3.576 & 1.176 & 3.587 \\
&
& (0.286) & (0.577) & (0.277) & (0.578) & (0.163) & (0.403) & (0.140) & (0.381) \\
& \multirow{2}{*}{POOL}
& 2.872 & 4.175 & 2.962 & 4.350 & 2.158 & 3.670 & 2.096 & 3.795 \\
&
& (0.572) & (0.638) & (0.654) & (0.631) & (0.282) & (0.404) & (0.302) & (0.386) \\
& \multirow{2}{*}{CNN}
& 1.453 & \textbf{3.499} & 1.466 & \textbf{3.580} & 1.336 & \textbf{3.467} & 1.349 & 3.543 \\
&
& (0.268) & \textbf{(0.600)} & (0.303) & \textbf{(0.576)} & (0.184) & \textbf{(0.416)} & (0.227) & (0.389) \\
& \multirow{2}{*}{HPS}
& 1.509 & 3.574 & 1.527 & 3.668 & 1.321 & 3.521 & 1.379 & 3.586 \\
&
& (0.298) & (0.609) & (0.316) & (0.577) & (0.205) & (0.425) & (0.222) & (0.390) \\
& \multirow{2}{*}{PRECAE}
& \textbf{1.133} & 3.684 & \textbf{1.120} & 3.647 & 1.131 & 3.631 & 1.117 & 3.591 \\
&
& \textbf{(0.248)} & (0.597) & \textbf{(0.236)} & (0.539) & (0.150) & (0.402) & (0.133) & (0.349) \\
& \multirow{2}{*}{PREMED}
& 1.272 & 4.119 & 1.258 & 4.101 & 1.277 & 4.006 & 1.246 & 4.106 \\
&
& (0.259) & (0.557) & (0.244) & (0.530) & (0.156) & (0.330) & (0.142) & (0.348) \\
& \multirow{2}{*}{PAIR}
& \textbf{1.161} & \textbf{3.552} & \textbf{1.161} & \textbf{3.629} & \textbf{1.096} & \textbf{3.478} & \textbf{1.089} & \textbf{3.527} \\
&
& \textbf{(0.253)} & \textbf{(0.580)} & \textbf{(0.253)} & \textbf{(0.556)} & \textbf{(0.161)} & \textbf{(0.409)} & \textbf{(0.157)} & \textbf{(0.379)} \\
\hline
\end{tabular}
}
\end{table}

Among the direct deep learning baselines, CNN tends to outperform HPS, especially for the CI group. A plausible explanation is that the CI group has a relatively larger sample size, so a directly trained CNN can already learn useful image-response patterns. Among the pretrained baselines, PRECAE generally performs better than PREMED, suggesting that, without more careful source-specific fine-tuning, representations learned directly from the hippocampus images through unsupervised reconstruction are more informative for MMSE prediction than generic embeddings extracted from a frozen foundation model. The strong performance of PRECAE in the 50\%-data CN setting is also reasonable, since its image representation is learned in a separate pretraining stage using all available images, rather than only the 50\% subset used in the downstream supervised fitting stage.

Compared to these baselines, PAIR is the only statistical method whose performance is consistently comparable to the deep learning methods. At the level of individual datasets, the prediction performance of PAIR is highly stable and is either the best or the second-best in all settings. This is particularly notable because PAIR is not a black-box prediction model, but instead explicitly models both shared structure and between-group heterogeneity. 
This conclusion becomes even clearer when overall performance across the two diagnostic groups is considered. To further support the above findings, Table~\ref{tab:adni_avg_perf} summarizes the average testing RMSE and the average group-wise testing correlation across the CN and CI groups under each setting. In terms of average RMSE, PAIR achieves the best performance in three of the four settings and is a very close second in the remaining one. A similar trend is observed for the average group-wise testing correlation, where PAIR attains the highest values under the full-data setting and remains highly competitive under the 50\%-data setting. Overall, these findings show that uncovering both shared patterns and heterogeneity in the imaging coefficients across diagnostic groups can enhance MMSE prediction, thereby providing a positive answer to (Q1).

\begin{table}[t!]
\renewcommand{\arraystretch}{0.7}
\centering
\caption{Average testing RMSE and average group-wise testing correlation across the CN and CI groups under the 50\%-data and full-data settings.}
\vskip -0.2in
\label{tab:adni_avg_perf}
\resizebox{\textwidth}{!}{
\begin{tabular}{cccccccccc}
\hline
~ & ~ & \multicolumn{4}{c}{50\% Data} & \multicolumn{4}{c}{Full Data} \\
\cline{3-10}
~ & Method & \multicolumn{2}{c}{Left} & \multicolumn{2}{c}{Right} & \multicolumn{2}{c}{Left} & \multicolumn{2}{c}{Right} \\
\cline{3-10}
~ & ~ & RMSE & Corr & RMSE & Corr & RMSE & Corr & RMSE & Corr \\
\hline
& SIRTV
& 2.562 & 0.131 & 2.595 & 0.136 & 2.382 & 0.157 & 2.381 & 0.177 \\
& CNN
& 2.476 & \textbf{0.168} & 2.523 & \textbf{0.164} & 2.401 & 0.173 & 2.446 & 0.165 \\
& HPS
& 2.542 & 0.129 & 2.598 & 0.142 & 2.421 & 0.131 & 2.482 & 0.144 \\
& PRECAE
& 2.409 & 0.116 & \textbf{2.383} & 0.150 & 2.381 & 0.143 & 2.354 & 0.177 \\
& PREMED
& 2.696 & 0.085 & 2.679 & 0.100 & 2.642 & 0.110 & 2.676 & 0.115 \\
& PAIR
& \textbf{2.357} & \textbf{0.163} & \textbf{2.395} & 0.149 & \textbf{2.287} & \textbf{0.195} & \textbf{2.308} & \textbf{0.191} \\
\hline
\end{tabular}
}
\end{table}

To further compare the practical efficiency of the proposed method and the deep learning baselines, we also examine their computation times. We do not include the two pretrained baselines PRECAE and PREMED in this comparison, since for these methods the dominant computational cost comes from the first-stage representation learning step. We thus focus on PAIR, CNN, and HPS. Table~\ref{tab:running_time} reports the combined running time (in seconds) for fitting both the CN and CI datasets within one fold. Across all settings, PAIR is consistently the fastest method, requiring roughly half the time of CNN and one-quarter the time of HPS. These results indicate that PAIR not only achieves strong predictive performance, but also enjoys a clear computational advantage over the best-performing deep learning baselines.

\begin{table}[t]
\renewcommand{\arraystretch}{0.7}
\centering
\caption{Combined running time (seconds) for fitting both CN and CI datasets within one fold.}
\vskip -0.2in
\label{tab:running_time}
\begin{tabular}{lcccc}
\hline
~ & \multicolumn{2}{c}{50\% Data} & \multicolumn{2}{c}{Full Data} \\
\cline{2-5}
Method & Left & Right & Left & Right \\
\hline
PAIR & 36.2 & 33.7 & 44.0 & 43.8 \\
CNN & 42.9 & 44.0 & 78.8 & 79.8 \\
HPS & 78.5 & 76.9 & 157.9 & 149.1 \\
\hline
\end{tabular}
\end{table}
\subsection{Shared and Specific Patterns Across Diagnostic Groups}

To address (Q2), we examine how baseline scalar covariates and hippocampal imaging contribute to MMSE prediction in the CN and CI diagnostic groups, and how these contributions differ across groups. Specifically, we compute the proportion of variance explained by (i) covariates alone, (ii) imaging after adjusting for covariates, and (iii) the joint model incorporating both covariates and imaging, using the training data from the CN and CI groups. Let $\widehat\beta_{t,Z}$ denote the ordinary least squares estimator for the $t$-th data source using only covariates $Z_t$, let $\bar Y_{t}=n_t^{-1}\sum_{i=1}^{n_t}Y_{ti}$ denote the mean of the response on the training data, and let $\{\widehat\beta_t, \widehat C_t\}$ denote the estimators obtained from our PAIR method. We define the predicted responses as $\widehat{Y}_{ti,Z} = \langle Z_{ti}, \widehat\beta_{t,Z} \rangle$ and $\widehat{Y}_{ti} = \langle Z_{ti}, \widehat\beta_{t} \rangle + \langle X_{ti}, \widehat{C}_t \rangle$ for the $t$-th data source. The explained variance ratios are defined as
\begin{align*}
    R_{t,Z}=\frac{\sum_{i=1}^{n_{t}}(\widehat Y_{ti,Z}-\bar Y_t)^2}{\sum_{i=1}^{n_{t}}(Y_{ti}-\bar Y_t)^2}, \qquad
    R_{t,X|Z}=\frac{\sum_{i=1}^{n_{t}}(\widehat Y_{ti}-\widehat Y_{ti,Z})^2}{\sum_{i=1}^{n_{t}}(Y_{ti}-\bar Y_t)^2}, \qquad
    R_{t,\{Z,X\}}=\frac{\sum_{i=1}^{n_{t}}(\widehat Y_{ti}-\bar Y_t)^2}{\sum_{i=1}^{n_{t}}(Y_{ti}-\bar Y_t)^2}.
\end{align*}

Table~\ref{tab:variance_explained} summarizes the mean values of these explained variance ratios across 500 runs. Several patterns emerge clearly. First, demographic covariates explain a nontrivial proportion of MMSE variation in both groups, with a stronger contribution in the CN group. In contrast, after adjusting for demographic covariates, hippocampal imaging contributes only modestly in the CN group but substantially more in the CI group. As a result, the joint contribution of covariates and imaging is also higher in the CI group than in the CN group. These findings indicate that hippocampal surfaces provide considerably more information about MMSE scores in the CI group than in the CN group once baseline demographic effects have been taken into account.
This pattern is clinically plausible. In CI patients, atrophy in the CA1 subregion of the hippocampus is often associated with disease severity \citep{frisoni2008mapping,de2015structural,li2024partially}. \citet{peng2015correlation} also found that baseline hippocampal volume was associated with MMSE scores in CI patients, whereas no such association was observed in healthy controls. Thus, in the CI group, hippocampal imaging more directly reflects structural degeneration relevant to cognitive decline, while in the CN group the additional information carried by imaging remains comparatively limited.

\begin{table}[t]
\renewcommand{\arraystretch}{0.7}
\centering
\caption{Mean variance explained (\%) by covariates, imaging after adjusting demographic covariates, and joint models for the CN and CI groups. The results are based on 500 runs.}
\vskip -0.2in
\label{tab:variance_explained}

\begin{tabular}{lcccccc}
\hline
~ & \multicolumn{3}{c}{Left} & \multicolumn{3}{c}{Right} \\
\cline{2-7}
Group & $R_{t,Z}$ & $R_{t,X|Z}$ & $R_{t,\{Z,X\}}$ & $R_{t,Z}$ & $R_{t,X|Z}$ & $R_{t,\{Z,X\}}$ \\
\hline
CN & 10.0\% & 3.2\% & 13.2\% & 10.0\% & 3.8\% & 13.8\% \\
CI & 5.9\% & 12.7\% & 18.7\% & 5.9\% & 11.7\% & 17.6\% \\
\hline
\end{tabular}
\end{table}

\begin{figure}[t!]
    \centering
    \includegraphics[width=\linewidth, height=0.65\textheight]{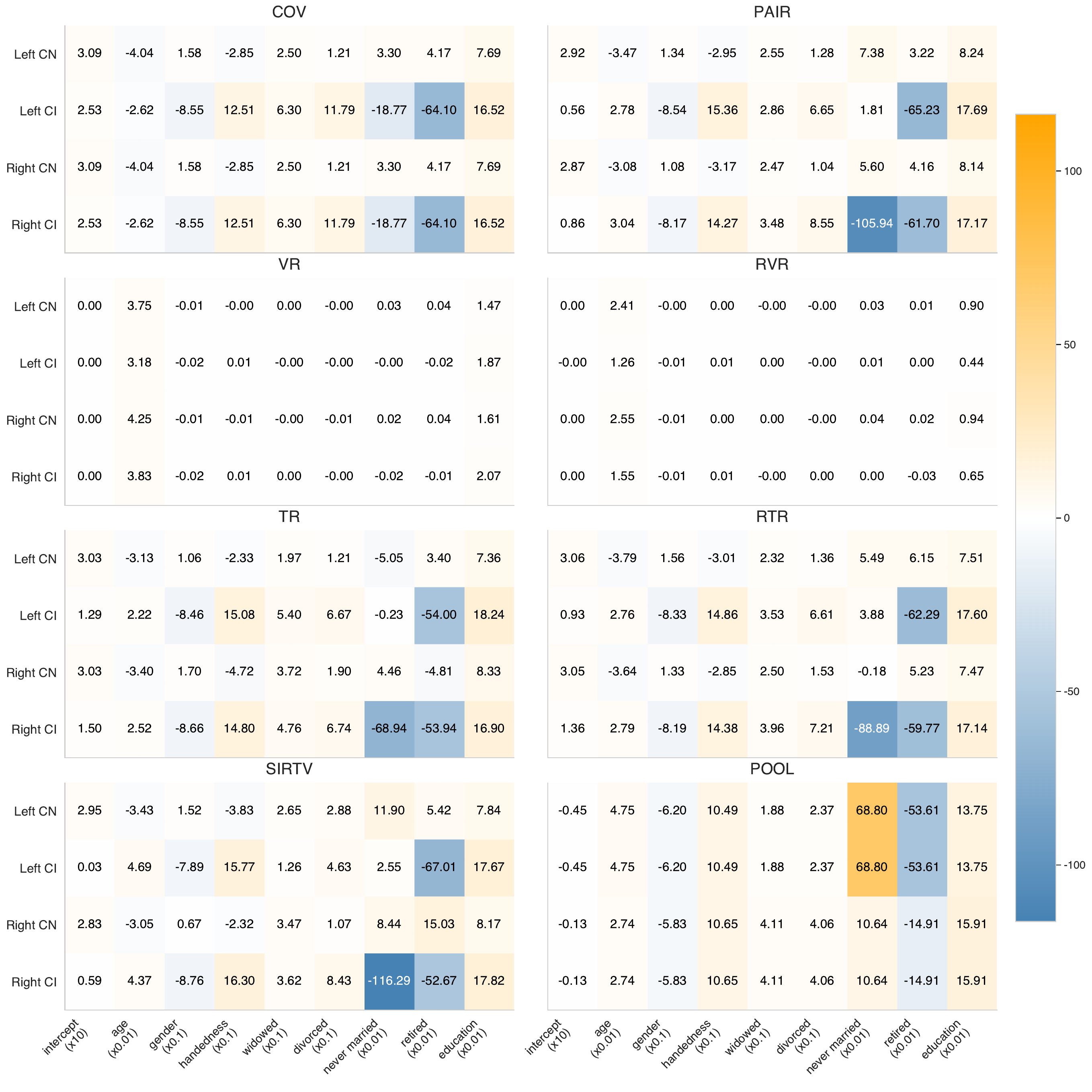}
    \vskip -0.2in
    \caption{Heatmaps of scaled mean covariate coefficients across different methods. Each coefficient is rescaled with the scaling factor shown on the horizontal axis.}
    \label{fig:covariate}
\end{figure}

To further understand how scalar effects vary across groups, Figure~\ref{fig:covariate} displays the scaled heatmaps of covariate coefficients across methods for both left and right hippocampal analyses. Clear heterogeneity is already visible when only demographic covariates are used. For example, the estimated signs of the coefficients for gender and handedness differ between the CN and CI groups, which is consistent with the marginal trends shown in Figure~\ref{fig:marginal correlations}. These findings are also broadly compatible with previous studies suggesting greater resilience to brain aging among cognitively healthy females, but more rapid decline among women after the onset of AD \citep{bourzac2025women}. By contrast, existing evidence suggests no significant differences between left- and right-handed individuals in language or visuospatial task performance \citep{abuduaini2024handedness}, so the handedness coefficients should be interpreted more cautiously.

The estimates from PAIR help clarify how imaging modifies these scalar associations. In the CN group, the covariate coefficients obtained by PAIR remain broadly consistent with those from the covariate-only model, which agrees with the small additional contribution of imaging seen in Table~\ref{tab:variance_explained}. In other words, once imaging is added, the estimated scalar effects in the CN group change only slightly. In the CI group, however, several coefficients shift noticeably after adjusting for hippocampal imaging. For instance, the coefficient for age changes from negative in the covariate-only model to positive after imaging is incorporated; a similar pattern also appears for the TR, RTR, and SIRTV methods. This suggests that, in the CI group, the relationship between age and cognition may be mediated or confounded by structural hippocampal degeneration, as aging is known to be associated with structural and functional changes in the hippocampus \citep{hu2020age}.

Additionally, we observe substantial changes in the coefficients for the intercept and the marital-status indicators (widowed, divorced, and never married) in the CI group after including imaging features. Notably, the signs of these coefficients remain broadly stable, suggesting that, relative to married individuals, widowed and divorced elderly participants may exhibit better cognitive performance, a pattern that is broadly consistent with prior studies \citep{hanes2024cognitive,karakose2025marital}. Compared with the competing methods, PAIR yields coefficient patterns that are both interpretable and well aligned with Table~\ref{tab:variance_explained}. VR and RVR produce coefficients that are close to zero for most scalar variables, while POOL ignores heterogeneity and therefore deviates substantially from the covariate-only patterns. TR, RTR, and SIRTV recover trends similar to those of PAIR.

\begin{figure}[t!]
    \centering
    \includegraphics[width=\linewidth, height=0.45\textheight]{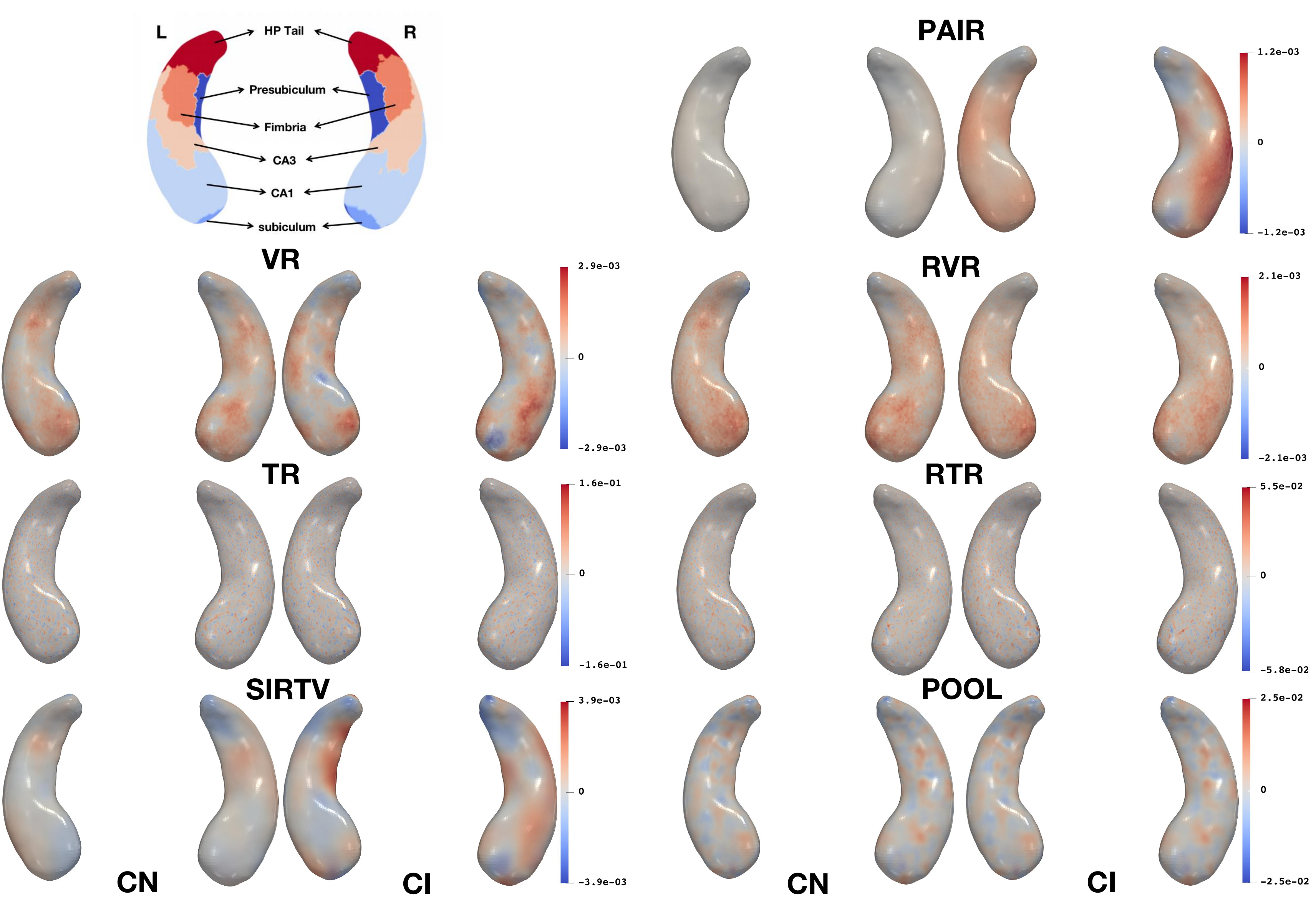}
    \vskip -0.2in
    \caption{Illustration of left and right hippocampus surface subregions and estimated left and right hippocampus surfaces for the CN and CI groups using different methods.}
    \label{fig:PAIR hamati}
\end{figure}

Figure~\ref{fig:PAIR hamati} further displays the estimated hippocampal surface effects associated with MMSE, averaged over 500 runs. Again, strong heterogeneity is observed between the CN and CI groups. In the CN group, the estimated signals on both left and right hippocampal surfaces are weak overall, and no clearly dominant subregions emerge. This is consistent with Table~\ref{tab:variance_explained}, which showed that hippocampal imaging adds only limited information after accounting for baseline demographic covariates in the CN group.

In contrast, the CI group exhibits pronounced positive surface signals on both hippocampi. The strongest effects are concentrated in subregions such as CA1, CA3, and the presubiculum, which is consistent with the literature showing that AD-related structural damage often begins in CA1 and then extends to other hippocampal subfields \citep{frisoni2008mapping,de2015structural,li2024partially}. Competing methods such as VR, RVR, and SIRTV reveal a broadly similar pattern in the CI group. However, PAIR produces more stable estimates across runs, as evidenced by the smaller variation in the estimated surfaces; see  Supplementary Material~\ref{supp: adni std}. Moreover, the CN--CI heterogeneity identified by PAIR in both CA1 and the presubiculum is noticeably stronger and more pronounced than that captured by SIRTV. This suggests that PAIR more effectively reveals the shared and group-specific roles of hippocampal structures across diagnostic groups. This finding is also supported by \citet{peng2015correlation}, who reported a strong association between hippocampal volume and MMSE scores in CI patients, but not in CN controls.
These results provide a direct answer to (Q2).  Scalar covariates contribute in both groups, whereas hippocampal imaging is substantially more informative in the CI group, and PAIR effectively reveals both shared and group-specific effect patterns in the scalar and imaging associations.
\section{Discussion}\label{sec: diss}
We developed the PAIR framework for integrating heterogeneous imaging and covariate data across diagnostic groups and applied it to map biological pathways linking hippocampal structure to cognitive decline in the ADNI study. Theoretically, we establish minimax-optimal error bounds that dynamically adapt to varying sharing structures. Numerically, our approach outperforms statistical baselines and achieves predictive accuracy comparable to advanced deep learning models, while providing superior interpretability, notably identifying established AD-related patterns in hippocampal subregions.

Several directions merit further investigation. First, the current framework assumes a bilinear representation of imaging coefficients; extending to nonlinear spatial components, for instance through neural network-based spatial encoders, could capture more complex structures while retaining interpretability. Second, applying PAIR to longitudinal imaging data would enable tracking of how brain-cognition associations evolve with disease progression, going beyond the cross-sectional analysis presented here. Finally, incorporating genetic and proteomic biomarkers within the multi-modality integration framework offers a promising avenue for building more comprehensive predictive models of AD.

{\bibhang=1.7pc
\renewcommand\bibname{\large \bf References}
\fontsize{12pt}{10.5pt plus.10pt minus .7pt}\selectfont
\setlength{\bibsep}{1pt}
\bibliographystyle{apalike}
\bibliography{PAIR}
}

\newpage
\appendix
\noindent The Supplementary Materials are organized as follows:
\begin{itemize}
    \item Supplementary Material~\ref{supp: simu res} presents additional simulation results, along with comprehensive implementation details for all evaluated methods.
    \item Supplementary Material~\ref{sec: supp adni analysis} provides further empirical results from the ADNI real data analysis, alongside the hyperparameter sensitivity analysis and specific implementation setups.
    \item Supplementary Material~\ref{sec:combat_harmonization} discusses the potential integration of PAIR with ComBat harmonization for addressing site/scanner heterogeneity.
    \item Supplementary Material~\ref{supp: proofs of theorem} contains the detailed proofs for the theoretical guarantees established in Section~\ref{sec:pair-theory}, including all supporting technical lemmas.
\end{itemize}

\section{Supplementary Simulation Results}\label{supp: simu res}
\subsection{Supplementary Visualization}\label{supp: simu data}
\begin{figure}[h]
    \centering
    \includegraphics[width=0.8\linewidth]{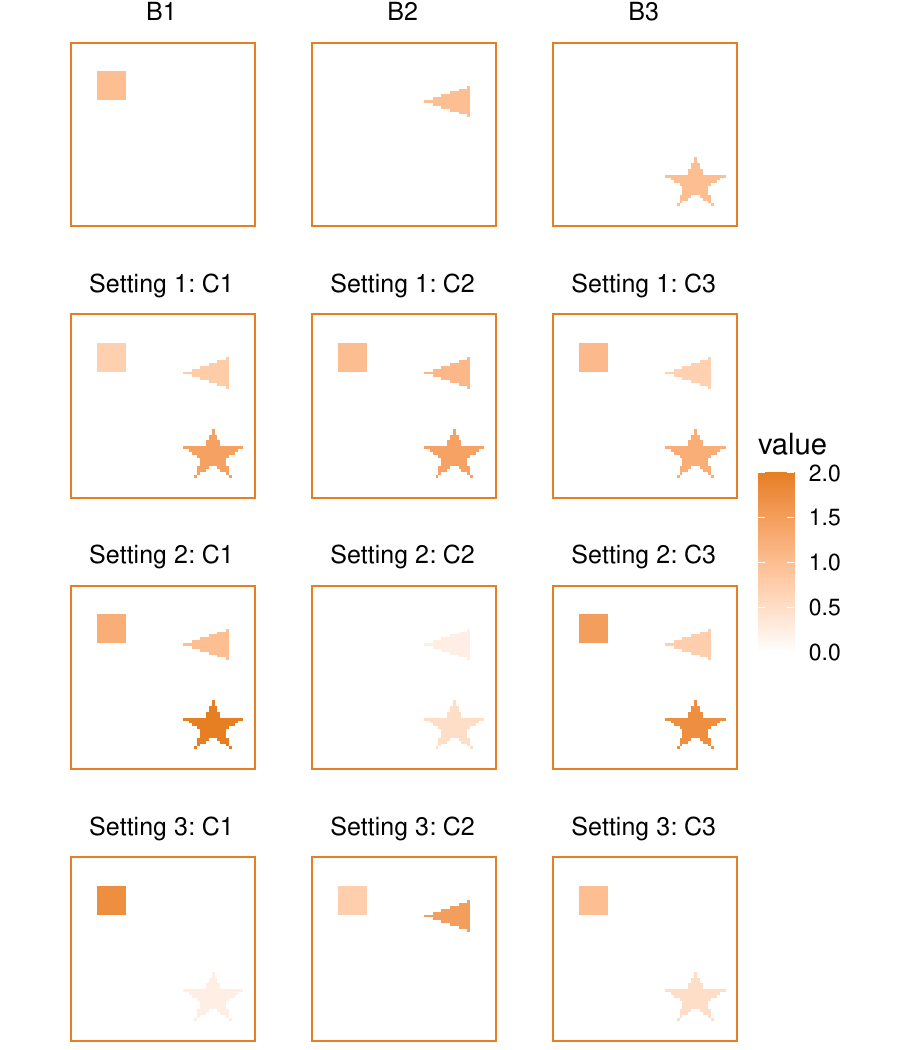}
    \vskip -0.2in
    \caption{Spatial component set and three imaging coefficients under three settings.}
    \label{fig:simu true coefficients}
\end{figure}

\begin{figure}
    \centering \includegraphics[width=1\linewidth]{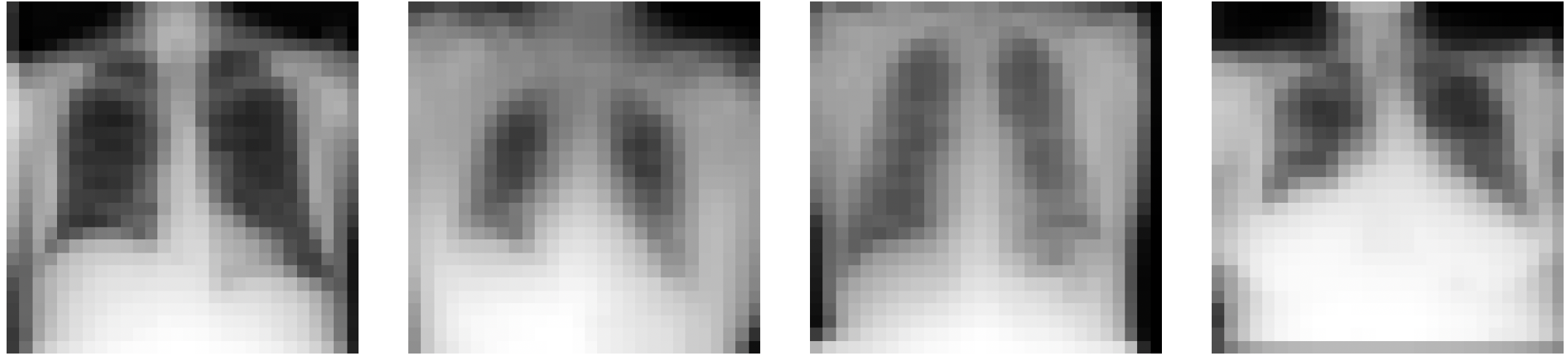}
    \vskip -0.2in
    \caption{Observed ChestMNIST images.}
    \label{fig:chestcase}
\end{figure}

Figure~\ref{fig:simu true coefficients} illustrates an example of the true coefficient structures under the three simulation settings described in Section~\ref{sec: simu}.
Figure~\ref{fig:chestcase} 
displays some observed ChestMNIST images. 

\subsection{Simulation based on Simulated Data}\label{sec:simu_simulated}
For this first simulation study, we generate the entries of the imaging covariates $X_{ti}$ independently from the standard normal distribution. For the sample size, we take $n_1=n_2=n_3=n\in\{300,600\}$. The estimation and prediction results are summarized in Table~\ref{tab:simu_all_metrics}. Since the three deep learning baselines do not directly estimate the imaging coefficients or regression coefficients, they are only compared in terms of the prediction metric ARMSE.

From Table~\ref{tab:simu_all_metrics}, larger sample sizes generally improve both estimation and prediction. The competing methods fall into three tiers. First, the noisy estimators VR, RVR, TR, and RTR all produce large estimation errors because they fail to capture the spatial structure: VR and RVR vectorize and destroy it, while TR and RTR rely on a CP decomposition that is too inflexible. Second, the spatially regularized methods SIRTV and POOL perform markedly better, though their relative ranking depends on heterogeneity---POOL can surpass SIRTV in Setting 1 but falls behind in Settings 2 and 3 as pooling becomes less appropriate. Third, PAIR achieves the best overall estimation and prediction across all three settings, with its advantage most pronounced in the small-sample regime; although its margin over SIRTV narrows as heterogeneity increases from Setting 1 to Setting 3, it remains the strongest method throughout. The deep learning baselines CNN, HPS, and PRE are not competitive here, likely because the sample size is still limited for fitting deep networks to structured spatial coefficients.

\begin{table}
\renewcommand{\arraystretch}{0.7}
\scriptsize
\setlength{\tabcolsep}{2.5pt}
\centering
\caption{Comparison of AEE$_C$, AEE$_\beta$, and ARMSE across various methods and simulated datasets under three settings with different sample sizes. The results are summarized by mean and standard deviation over 100 repeats.}
\vskip -0.2in
\label{tab:simu_all_metrics}
\resizebox{\textwidth}{!}{
\begin{tabular}{lllcccccccccc}
\hline
Setting & $n$ & Metric & VR & RVR & TR & RTR & SIRTV & POOL & PAIR & CNN & HPS & PRE \\
\hline
\multirow{12}{*}{Setting 1}
& \multirow{6}{*}{300}
& \multirow{2}{*}{AEE$_C$}
& 17.75 & 17.87 & 30.20 & 18.55 & 7.92 & 5.90 & \textbf{1.86} & -- & -- & -- \\
&& & (1.82) & (1.83) & (3.09) & (1.93) & (1.27) & (1.30) & \textbf{(0.47)} & -- & -- & -- \\
&& \multirow{2}{*}{AEE$_{\beta}$}
& 2.01 & 1.95 & 4.83 & 3.01 & 1.30 & 1.78 & \textbf{0.35} & -- & -- & -- \\
&& & (0.40) & (0.39) & (1.03) & (0.59) & (0.31) & (0.41) & \textbf{(0.09)} & -- & -- & -- \\
&& \multirow{2}{*}{ARMSE}
& 17.88 & 17.96 & 30.30 & 18.84 & 8.06 & 6.25 & \textbf{2.14} & 14.52 & 18.05 & 18.49 \\
&& & (2.00) & (1.99) & (3.46) & (2.10) & (1.27) & (1.20) & \textbf{(0.43)} & (1.56) & (2.03) & (2.06) \\
\cline{2-13}
& \multirow{6}{*}{600}
& \multirow{2}{*}{AEE$_C$}
& 17.35 & 17.35 & 18.48 & 14.14 & 1.88 & 4.98 & \textbf{0.97} & -- & -- & -- \\
&& & (1.77) & (1.77) & (2.89) & (2.47) & (0.19) & (1.23) & \textbf{(0.19)} & -- & -- & -- \\
&& \multirow{2}{*}{AEE$_{\beta}$}
& 1.93 & 1.77 & 2.07 & 1.58 & 0.24 & 1.74 & \textbf{0.16} & -- & -- & -- \\
&& & (0.35) & (0.32) & (0.50) & (0.40) & (0.05) & (0.38) & \textbf{(0.03)} & -- & -- & -- \\
&& \multirow{2}{*}{ARMSE}
& 17.48 & 17.44 & 18.66 & 14.24 & 2.15 & 5.42 & \textbf{1.41} & 13.03 & 17.66 & 18.39 \\
&& & (1.82) & (1.84) & (2.98) & (2.52) & (0.20) & (1.17) & \textbf{(0.17)} & (1.39) & (1.94) & (1.97) \\
\hline
\multirow{12}{*}{Setting 2}
& \multirow{6}{*}{300}
& \multirow{2}{*}{AEE$_C$}
& 19.88 & 19.97 & 32.62 & 18.74 & 7.61 & 11.63 & \textbf{2.80} & -- & -- & -- \\
&& & (0.82) & (0.82) & (1.95) & (2.05) & (1.35) & (1.82) & \textbf{(1.01)} & -- & -- & -- \\
&& \multirow{2}{*}{AEE$_{\beta}$}
& 2.08 & 1.99 & 5.23 & 2.96 & 1.23 & 2.12 & \textbf{0.49} & -- & -- & -- \\
&& & (0.37) & (0.34) & (1.05) & (0.68) & (0.32) & (0.41) & \textbf{(0.17)} & -- & -- & -- \\
&& \multirow{2}{*}{ARMSE}
& 19.95 & 20.06 & 32.79 & 18.99 & 7.78 & 11.90 & \textbf{3.05} & 16.28 & 20.29 & 20.69 \\
&& & (1.31) & (1.31) & (2.46) & (2.33) & (1.40) & (1.97) & \textbf{(1.01)} & (1.08) & (1.31) & (1.35) \\
\cline{2-13}
& \multirow{6}{*}{600}
& \multirow{2}{*}{AEE$_C$}
& 19.43 & 19.23 & 18.28 & 12.90 & 2.10 & 11.11 & \textbf{1.09} & -- & -- & -- \\
&& & (0.80) & (0.79) & (3.03) & (2.09) & (0.59) & (1.89) & \textbf{(0.15)} & -- & -- & -- \\
&& \multirow{2}{*}{AEE$_{\beta}$}
& 1.92 & 1.69 & 2.04 & 1.42 & 0.26 & 1.86 & \textbf{0.17} & -- & -- & -- \\
&& & (0.35) & (0.33) & (0.50) & (0.36) & (0.09) & (0.35) & \textbf{(0.04)} & -- & -- & -- \\
&& \multirow{2}{*}{ARMSE}
& 19.48 & 19.26 & 18.42 & 12.93 & 2.38 & 11.36 & \textbf{1.51} & 14.37 & 19.64 & 20.52 \\
&& & (1.11) & (1.10) & (3.25) & (2.10) & (0.55) & (1.92) & \textbf{(0.13)} & (0.86) & (1.16) & (1.11) \\
\hline
\multirow{12}{*}{Setting 3}
& \multirow{6}{*}{300}
& \multirow{2}{*}{AEE$_C$}
& 15.42 & 15.47 & 26.46 & 13.40 & 4.29 & 11.00 & \textbf{2.58} & -- & -- & -- \\
&& & (2.44) & (2.45) & (3.92) & (3.14) & (1.66) & (2.35) & \textbf{(1.17)} & -- & -- & -- \\
&& \multirow{2}{*}{AEE$_{\beta}$}
& 2.00 & 1.84 & 4.22 & 2.23 & 0.74 & 2.03 & \textbf{0.46} & -- & -- & -- \\
&& & (0.38) & (0.35) & (1.04) & (0.72) & (0.32) & (0.40) & \textbf{(0.21)} & -- & -- & -- \\
&& \multirow{2}{*}{ARMSE}
& 15.65 & 15.64 & 26.81 & 13.69 & 4.53 & 11.29 & \textbf{2.88} & 12.55 & 15.92 & 16.14 \\
&& & (2.49) & (2.51) & (4.16) & (3.30) & (1.62) & (2.46) & \textbf{(1.10)} & (2.02) & (2.60) & (2.58) \\
\cline{2-13}
& \multirow{6}{*}{600}
& \multirow{2}{*}{AEE$_C$}
& 15.07 & 14.71 & 11.56 & 8.16 & 1.15 & 10.66 & \textbf{1.05} & -- & -- & -- \\
&& & (2.37) & (2.34) & (3.67) & (2.39) & (0.30) & (2.29) & \textbf{(0.20)} & -- & -- & -- \\
&& \multirow{2}{*}{AEE$_{\beta}$}
& 1.96 & 1.52 & 1.32 & 0.93 & 0.18 & 1.89 & \textbf{0.17} & -- & -- & -- \\
&& & (0.41) & (0.31) & (0.56) & (0.36) & (0.05) & (0.41) & \textbf{(0.03)} & -- & -- & -- \\
&& \multirow{2}{*}{ARMSE}
& 15.16 & 14.70 & 11.63 & 8.29 & 1.57 & 10.87 & \textbf{1.48} & 11.11 & 15.48 & 15.94 \\
&& & (2.28) & (2.31) & (3.66) & (2.35) & (0.25) & (2.21) & \textbf{(0.13)} & (1.81) & (2.46) & (2.45) \\
\hline
\end{tabular}
}
\end{table}

\begin{figure}
    \centering
    \includegraphics[width=0.8\linewidth]{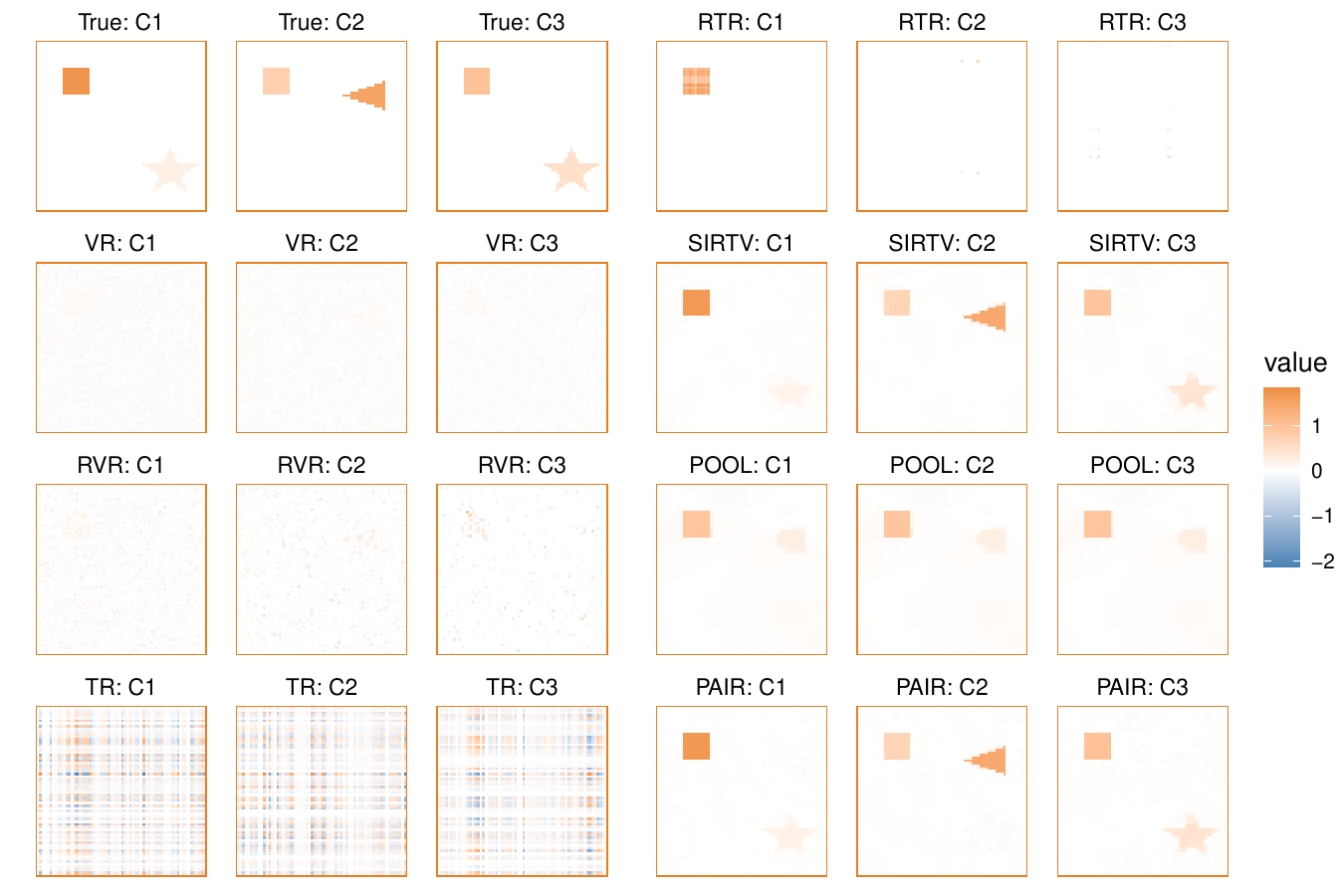}
    \vskip -0.2in
    \caption{True imaging coefficients and the estimated coefficients using different methods and simulated data with sample size $n_1=n_2=n_3=300$.}
    \label{fig:simulation}
\end{figure}

Figure~\ref{fig:simulation} visualizes the true and estimated imaging coefficients for one experiment in Setting 3. The true coefficients follow a partially-shared structure: a square signal shared by all three sources, a pentagon shared by $C_1$ and $C_3$, and a source-specific triangle in $C_2$. The noisy estimators (VR, RVR, TR, RTR) can only locate signals roughly; CP-based TR and RTR recover the simple square but miss the nonlinear pentagon and triangle. The spatially regularized methods have limitations too: SIRTV recovers the square but struggles with the complex pentagon at this sample size, and POOL introduces substantial noise by ignoring heterogeneity. PAIR recovers all three signal components accurately, reflecting its ability to borrow strength across sources while respecting source-specific structure.

\subsection{Implementation Details}
\label{sec:simu-implementation}
\noindent\textbf{Implementation Details for Simulation based on Simulated Data}.
To complement the statistical baselines in Section~\ref{sec:simu_simulated}, we additionally consider three deep learning baselines, denoted by CNN, HPS, and PRE. Let $(Y_{ti},Z_{ti},X_{ti})$ be as in Section~\ref{sec: simu} with $p=q=64$. All three methods directly predict $Y_{ti}$ from $(X_{ti}, Z_{ti})$, and are trained separately on the simulated training, validation, and testing splits used in the main experiments.

\textbf{CNN.}
For the CNN baseline, each image is reshaped into a single-channel tensor of size $1\times p\times q$. The image encoder contains three convolutional blocks: $\text{Conv}(1,16,5\times 5)\rightarrow \text{ReLU}\rightarrow \text{MaxPool}(2)$, $\text{Conv}(16,32,3\times 3)\rightarrow \text{ReLU}\rightarrow \text{MaxPool}(2)$, and $\text{Conv}(32,64,3\times 3)\rightarrow \text{ReLU}\rightarrow \text{AdaptiveAvgPool}(4,4)$. The resulting feature map is flattened into a $64\times 4\times 4=1024$ dimensional image representation, which is concatenated with the covariate vector $Z_{ti}$ and passed through a prediction head $\mathbb{R}^{1024+d}\rightarrow \mathbb{R}^{128}\rightarrow \mathbb{R}$ with ReLU activation and dropout.

\textbf{HPS.}
The HPS baseline combines a global branch and a patch-summary branch. The global branch applies $\text{Conv}(1,16,3\times 3)\rightarrow \text{ReLU}\rightarrow \text{MaxPool}(2)\rightarrow \text{Conv}(16,32,3\times 3)\rightarrow \text{ReLU}\rightarrow \text{AdaptiveAvgPool}(4,4)$, followed by a linear projection into a $d_e$-dimensional embedding. In the reported simulation experiments, $d_e$ is fixed at $64$. The patch-summary branch first adaptively pools the image to $32\times 32$, partitions it into non-overlapping $8\times 8$ patches, maps each patch into $\mathbb{R}^{d_e}$, and then aggregates the patch embeddings through an attention multilayer perceptron followed by softmax weighting. The global embedding, the patch-summary embedding, and the covariates are concatenated and fed into a prediction head $\mathbb{R}^{2d_e+d}\rightarrow \mathbb{R}^{128}\rightarrow \mathbb{R}$ with ReLU activation and dropout.

\textbf{PRE.}
The third baseline, denoted by PRE, uses a pretrained image encoder followed by a supervised prediction head. Specifically, we adopt a ResNet18 backbone, modify the first convolutional layer to accept a single-channel image input, and replace the final classification layer by an identity mapping so that the backbone produces a $512$-dimensional image representation. This representation is concatenated with $Z_{ti}$ and passed through a prediction head $\mathbb{R}^{512+d}\rightarrow \mathbb{R}^{128}\rightarrow \mathbb{R}$. In the reported simulation experiments, the pretrained backbone is kept frozen and only the prediction head is trained.

\textbf{Training and model selection.}
For all three methods, the loss function is the mean squared prediction error on the training split. Optimization is carried out using Adam with gradient clipping, and model selection is based on the validation mean squared error. Across all three simulation settings, the maximum number of epochs is set to $1000$ and early stopping with patience $50$ is used. The hidden dimension in the prediction head is fixed at $128$, the batch size is selected from $\{32,64\}$, and the weight decay is selected from $\{0,10^{-4}\}$. The setting-specific learning-rate and dropout grids are listed in Table~\ref{tab:simu_dl_grids}. For HPS, the patch size and pooled image size are fixed at $8$ and $32\times 32$, respectively, and for PRE the pretrained backbone remains frozen throughout the simulation experiments.

\begin{table}[h]
\centering
\small
\caption{Setting-specific learning-rate and dropout grids for the CNN, HPS, and PRE baselines in the fully simulated and ChestMNIST simulations.}
\label{tab:simu_dl_grids}
\begin{tabular}{lll}
\hline
Setting   & Learning rate                                        & Dropout            \\
\hline
Setting 1 & $\{10^{-3}, 3\times 10^{-4}, 10^{-4}\}$              & $\{0, 0.2\}$        \\
Setting 2 & $\{10^{-3}, 10^{-4}, 3\times 10^{-5}\}$              & $\{0, 0.2, 0.4\}$   \\
Setting 3 & $\{3\times 10^{-4}, 10^{-4}, 3\times 10^{-5}\}$      & $\{0.2, 0.4\}$      \\
\hline
\end{tabular}
\end{table}

\textbf{Tuning ranges for statistical baselines.}
For the remaining methods, the tuning ranges are largely shared across the three simulation settings. Specifically, RVR selects the $\ell_1$ penalty from $\{0.01, 0.1, 1, 10, 100\}$, TR selects the rank from $\{1,2,3,4,5\}$, and RTR tunes both the rank and the $\ell_1$ penalty over $\{1,2,3,4,5\}\times\{0.01, 0.1, 1, 10, 100\}$. For SIRTV and POOL, the TV penalty is selected from $\{0.01, 0.1, 1, 10, 100\}$. For PAIR, the candidate grids in Settings 1 and 2 are $R\in\{2,3,4\}$, $\lambda\in\{0.01, 0.1, 1, 10\}$, $\gamma\in\{0.01, 0.1, 1, 10\}$, and $\tau\in\{0.5,0.7\}$, while Setting 3 uses the same grids except that $\tau$ is expanded to $\{0.3,0.5,0.7\}$. In terms of fixed optimization settings, VR is trained by SGD with learning rate $0.01$ for $1000$ epochs. RVR uses SGD with learning rate $0.01$, maximum $3000$ epochs, and patience $1000$. TR and RTR use Adam with learning rate $0.01$ for the low-rank factors, with maximum $3000$ epochs and patience $1000$. SIRTV and POOL use Adam with learning rate $0.01$, maximum $1000$ epochs, and patience $100$. PAIR uses Adam with learning rates $0.01$ for both $B$ and $W$, maximum $3000$ epochs, and patience $100$.

\noindent\textbf{Implementation Details for Simulation based on ChestMNIST Data}.
The architectures, optimizer, training schedule, model-selection procedure, and tuning grids (for both deep-learning and statistical baselines) are identical to those in Section~\ref{sec:simu_simulated}, except that (i) the image covariates $X_{ti}$ are drawn from the ChestMNIST training split and standardized before use (as described in Section~\ref{sec: simu}), and (ii) for PRE, the final residual block of the ResNet18 backbone is kept trainable so that the representation can better adapt to ChestMNIST images.
\section{Supplementary ADNI Data Analysis Results}\label{sec: supp adni analysis}
\subsection{Supplementary Standard Deviation Results for Imaging Coefficients}\label{supp: adni std}
\begin{figure}[htbp]
    \centering
    \includegraphics[width=1\linewidth]{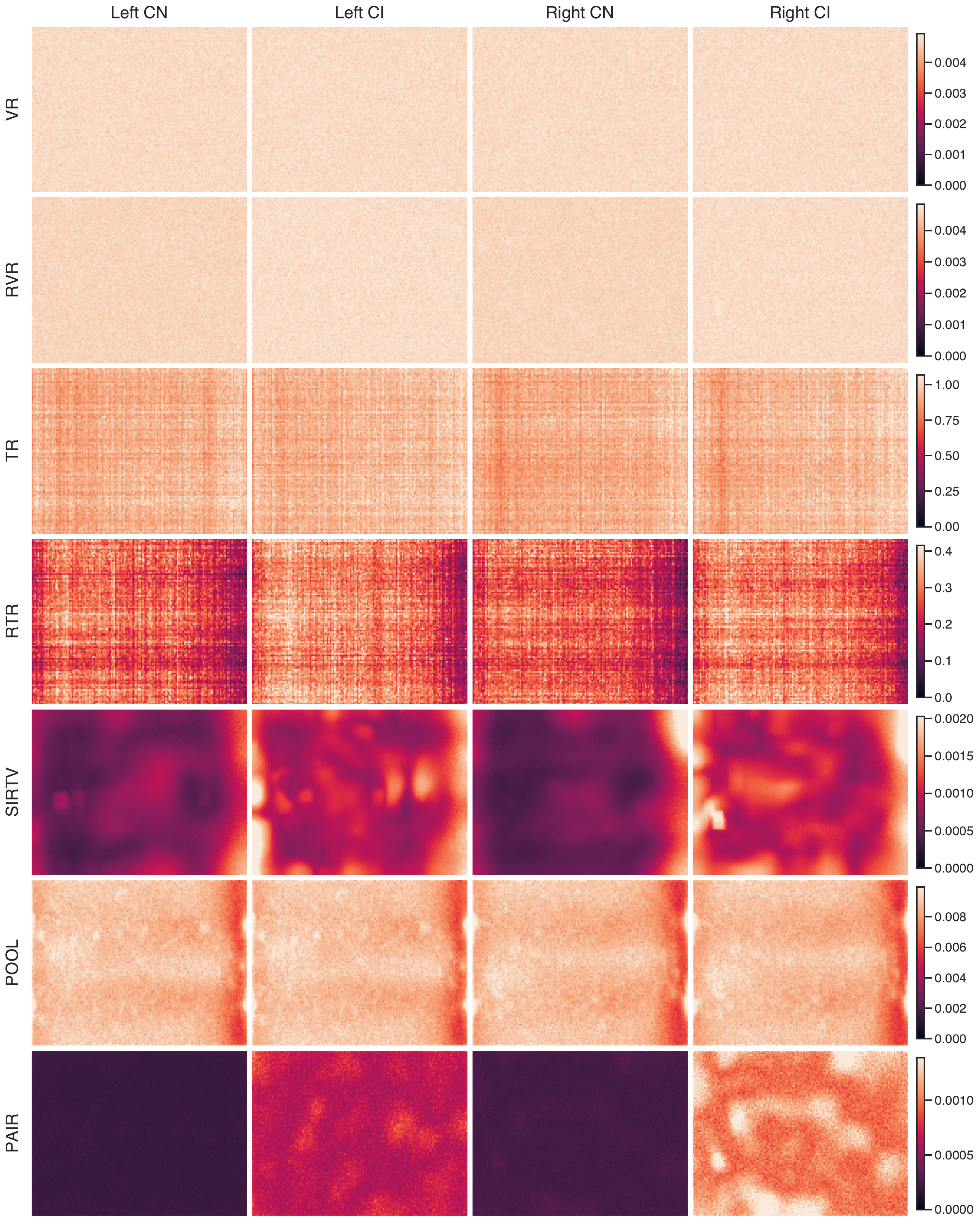}
    \vskip -0.2in
    \caption{Empirical standard deviations of the estimated imaging coefficients on the left and right hippocampus surfaces for the CN and CI groups, compared across different methods over 500 runs.}
    \label{fig: adni std} 
\end{figure}

Figure~\ref{fig: adni std} presents the empirical standard deviations of the estimated imaging coefficients for the left and right hippocampus surfaces across the CN and CI groups. As observed, our proposed PAIR yields smaller standard deviations compared to the alternative methods, indicating a highly stable estimation.

\subsection{Hyperparameter Sensitivity Analysis}
\label{sec:sensitivity_analysis}

\begin{figure}[htbp]
    \centering
    \includegraphics[width=1\linewidth]{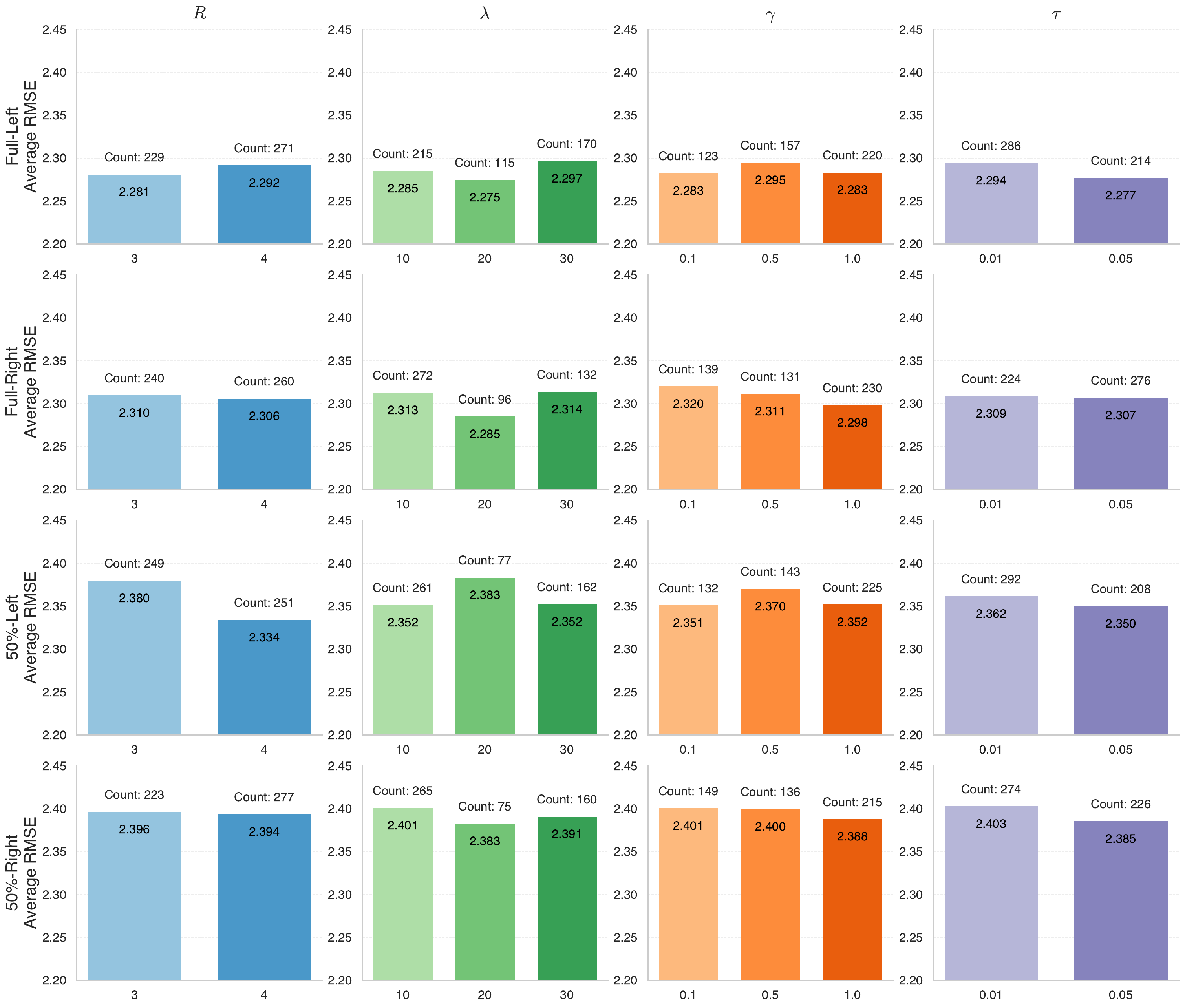}
    \caption{Hyperparameter selection frequencies and the corresponding mean ARMSE over 500 runs for the PAIR model. The four panels represent the four experimental settings (Left/Right hippocampus with Full/50\% data). Each subplot illustrates the marginal effect of tuning $R$, $\lambda$, $\gamma$, and $\tau$, respectively.}
    \label{fig:pair_tuning_rmse}
\end{figure}

The proposed PAIR model involves four tuning parameters: the rank $R$, the TV penalty $\lambda$, the rank penalty $\gamma$, and the threshold $\tau$. To evaluate the sensitivity of PAIR to these hyperparameters, we independently analyze the selection frequency and the resulting predictive performance for each parameter across 500 runs (comprising 100 repetitions of 5-fold cross-validation). Figure~\ref{fig:pair_tuning_rmse} summarizes the selection counts and the corresponding mean of ARMSE for each parameter value across the four experimental configurations (left and right hippocampus, with full and 50\% data). For instance, in the full data setting for the left hippocampus (first row, first column of Figure~\ref{fig:pair_tuning_rmse}), $R=3$ is selected 229 times yielding a mean ARMSE of 2.281, while $R=4$ is selected 271 times with a mean ARMSE of 2.292. The performance difference is practically negligible. Across all settings and all four hyperparameters, the mean ARMSE values remain remarkably consistent regardless of the specific parameter value chosen by the validation procedure. This indicates that the predictive accuracy of the PAIR framework is robust and minimally sensitive to hyperparameter selection within the candidate grid.

\begin{figure}[htbp]
    \centering
    \includegraphics[width=1\linewidth]{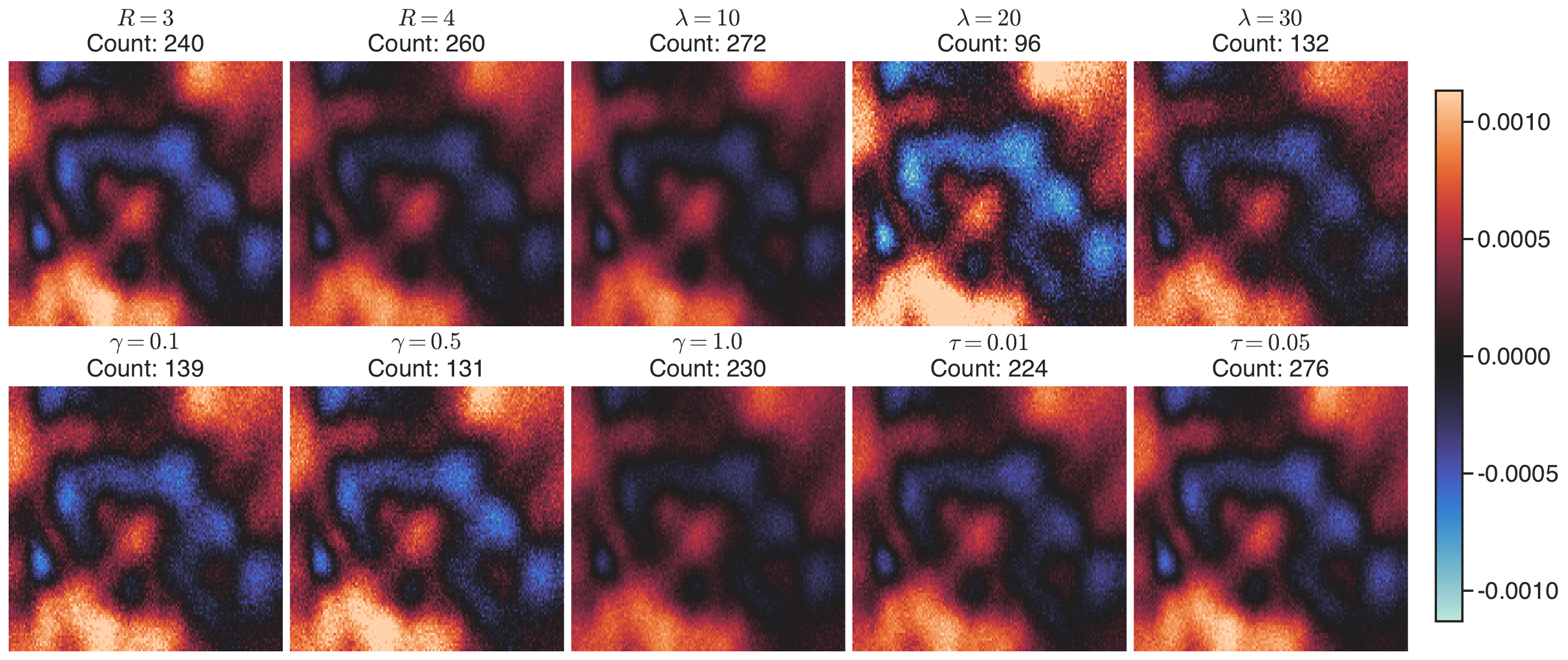}
    \caption{Mean estimated imaging coefficients for the right hippocampus in the CI group (full data setting), categorized by the selected hyperparameter values across the 500 runs. The structural patterns remain highly stable across different choices of $R$, $\lambda$, $\gamma$, and $\tau$.}
    \label{fig:pair_paramwise_coef}
\end{figure}

To further investigate whether different hyperparameter choices fundamentally alter the structural interpretation of the imaging coefficients, we visualize the estimated spatial patterns conditioned on the selected parameter values. Using the right hippocampus of the CI group under the full data setting as a representative example, Figure~\ref{fig:pair_paramwise_coef} displays the averaged imaging coefficients marginalized over the specific values of $R, \lambda, \gamma$, and $\tau$ chosen during the 500 runs. For example, the top two panels correspond to runs where $R=3$ (selected 240 times) and $R=4$ (selected 260 times). The estimated spatial patterns under both rank selections are visually highly consistent, preserving the identical active regions and signal intensities. Similar highly consistent spatial patterns are observed across the different candidate values for $\lambda$, $\gamma$, and $\tau$. These results strongly suggest that the estimation procedure of PAIR is stable.

\subsection{Implementation Details}
\label{sec:adni-dl-pretrained}
\noindent\textbf{Deep Learning and Pretrained Representation Baselines.}
To provide a comprehensive and fair comparison with modern predictive approaches, we additionally consider four baselines: CNN, HPS, PRECAE, and PREMED. CNN and HPS are direct image-covariate prediction models, whereas PRECAE and PREMED first construct image representations and subsequently combine them with demographic covariates in a downstream ridge regression model. Let $X_{ti}$ and $Z_{ti}$ be the hippocampal surface image and demographic covariates for subject $i$ in source $t$ as defined in Section~\ref{sec: real data}, with $X_{ti}\in\mathbb{R}^{100\times 150}$ and $Z_{ti}\in\mathbb{R}^{p_0}$.

\textbf{CNN.}
For the CNN baseline, the image $X_{ti}$ is first reshaped into a single-channel input tensor of size $1\times 100\times 150$. The image encoder consists of three convolutional blocks: $\text{Conv}(1,16,5\times 5)\rightarrow \text{ReLU}\rightarrow \text{MaxPool}(2)$, $\text{Conv}(16,32,3\times 3)\rightarrow \text{ReLU}\rightarrow \text{MaxPool}(2)$, and $\text{Conv}(32,64,3\times 3)\rightarrow \text{ReLU}\rightarrow \text{AdaptiveAvgPool}(4,4)$. The resulting image representation is flattened into a feature vector of dimension $64\times 4\times 4=1024$. This image feature is concatenated with the demographic covariate vector $Z_{ti}$ and passed through a fully connected prediction head $\mathbb{R}^{1024+p_0}\rightarrow \mathbb{R}^{128}\rightarrow \mathbb{R}$, utilizing a ReLU activation and a dropout layer between the two linear mappings.

\textbf{HPS.}
For the HPS method, the image representation is constructed from two complementary branches: a global branch and a patch-summary branch. The global branch applies $\text{Conv}(1,16,3\times 3)\rightarrow \text{ReLU}\rightarrow \text{MaxPool}(2)\rightarrow \text{Conv}(16,32,3\times 3)\rightarrow \text{ReLU}\rightarrow \text{AdaptiveAvgPool}(4,4)$, followed by a linear projection mapping the dimension from $32\times 4\times 4$ to an embedding vector in $\mathbb{R}^{d_e}$, where $d_e\in\{64,128\}$ is a tunable hyperparameter. The patch-summary branch first adaptively pools the image to a size of $32\times 32$, then partitions it into non-overlapping $8\times 8$ patches. Since the input is single-channel, each patch is represented by a $64$-dimensional vector, which is projected into $\mathbb{R}^{d_e}$ via a linear embedding layer. An attention multilayer perceptron is subsequently employed to produce a scalar score for each patch, followed by softmax normalization across patches. The final patch-level representation is obtained as the weighted average of all embedded patches. The global image embedding, the patch-summary embedding, and the demographic covariates are then concatenated and fed into a fully connected prediction head $\mathbb{R}^{2d_e+p_0}\rightarrow \mathbb{R}^{128}\rightarrow \mathbb{R}$, again utilizing ReLU activation and dropout.

\textbf{Training and model selection for CNN and HPS.}
For both CNN and HPS, the loss function is the mean squared prediction error on the training split. Optimization is carried out using Adam, with gradient clipping applied at a maximum norm of $5$. For each candidate hyperparameter configuration, the model is trained for at most $300$ epochs. Early stopping with a patience of $30$ is employed based on the validation mean squared error. The optimal hyperparameter configuration is selected via the validation error, and the chosen model is then evaluated on the testing split.

\textbf{PRECAE.}
The first pretrained baseline, denoted as PRECAE, utilizes a convolutional autoencoder (CAE). For each side, all hippocampal surface images are read as $100\times 150$ single-channel inputs. The images are standardized using the global mean and standard deviation estimated from a randomly selected training subset, with $15\%$ of the images held out as a validation subset for unsupervised model selection. The CAE is trained exclusively on image data and does not leverage MMSE labels during representation learning. The encoder comprises four strided convolutional blocks, progressively reducing the image resolution from $100\times 150$ to $50\times 75$, $25\times 38$, $13\times 19$, and finally $7\times 10$, while increasing channel sizes from $1$ to $c$, $2c$, $4c$, and $8c$, where $c$ is the base channel size. Each convolution layer is followed by batch normalization and a ReLU activation. The final feature map is flattened and mapped to a latent representation of dimension $d$ through a fully connected layer. The decoder is a mirrored transposed-convolution network that reconstructs the input image from the latent code. 

The candidate hyperparameter configurations are $(c,d,\mathrm{lr})\in\{(16,64,10^{-3}), (32,64,10^{-3})$, $ (32,128,10^{-3}), (32,128,3\times 10^{-4})\}$. For each configuration, the CAE is trained by minimizing the reconstruction mean squared error $\frac{1}{n}\sum_{i=1}^n \|\widehat X_i-X_i\|_F^2$ using Adam with a weight decay of $10^{-5}$. Mixed precision training is enabled to accelerate optimization. The maximum number of epochs is set to $80$, and early stopping with a patience of $12$ is applied according to the validation reconstruction loss. After identifying the best CAE configuration, the encoder extracts a fixed latent representation for every image. In the second stage, for each source and repeated cross-validation split, the extracted CAE representation $R_{ti}$ is concatenated with the demographic covariates to form the downstream feature vector $F_{ti}=(R_{ti}^\top,Z_{ti}^\top)^\top$. Responses are predicted via ridge regression, standardizing features using the training subset and centering the response. The ridge penalty parameter is selected from $\{10^{-4}, 10^{-3}, 10^{-2}, 10^{-1}, 1, 10, 100\}$ based on validation RMSE. This downstream ridge step remains consistent across both the full-data and 50\% settings.

\textbf{PREMED.}
The second pretrained baseline, PREMED, employs a frozen medical imaging foundation model to extract embeddings. We utilize MedImageInsight, a general-domain medical image embedding model~\citep{codella2024medimageinsight} (resources available at the \href{https://github.com/microsoft/healthcareai-examples}{Microsoft healthcare AI examples repository}). Specifically, we load the pretrained vision and language checkpoints without any fine-tuning, keeping the encoder entirely frozen. Consequently, tuning in PREMED focuses on image preprocessing strategies rather than network parameters. 

For each side, the image-only data is randomly partitioned into an $85\%$ training subset and a $15\%$ validation subset. Four candidate preprocessing strategies are evaluated: \textit{minmax} (rescales using minimum and maximum pixel values), \textit{pct\_1\_99} and \textit{pct\_2\_98} (clips to the respective percentile ranges before rescaling), and \textit{zclip\_3} (standardizes using training-subset statistics and clips to $[-3,3]$ before rescaling). Each processed image is converted to an 8-bit PNG and passed through MedImageInsight to yield an embedding. To determine the optimal preprocessing strategy, a proxy augmentation involving random intensity scaling and additive Gaussian noise is applied to the validation images. Let $E_i$ and $E_i^{\mathrm{aug}}$ denote the embeddings extracted from the original and augmented validation images, respectively. After row-wise $\ell_2$ normalization, the proxy score for model selection is defined as the average cosine similarity between the original and augmented embeddings, regularized by a small variance term to encourage feature diversity. The preprocessing strategy yielding the highest proxy score is selected, and the frozen encoder extracts the final embeddings for all images. Subsequently, PREMED employs the exact same downstream supervised ridge regression stage as PRECAE, utilizing the identical hyperparameter grid and validation procedure. For the 50\% setting, subsampling operates independently within each repeat prior to cross-validation.

\noindent\textbf{Tuning Ranges and Fixed Optimization Settings}
To ensure clarity and reproducibility, we summarize the hyperparameter search spaces and fixed optimization settings. Table~\ref{tab:adni_tuning_traditional} details the configurations for the traditional and statistical methods, while Table~\ref{tab:adni_tuning_dl} outlines the settings for the deep learning and pretrained representation baselines. 

The defined ranges are applied consistently across all four experimental settings (left hippocampus with full data, left hippocampus with 50\% data, right hippocampus with full data, and right hippocampus with 50\% data). Furthermore, for statistical methods that rely on gradient-based optimization, learning rates are held constant throughout the implementation and are therefore categorized under \emph{fixed settings} rather than \emph{tuned hyperparameters}.

% ==========================================
% Table 1: Traditional and Statistical Methods
% ==========================================
\begin{table}[htpb]
\centering
\small 
\caption{Hyperparameter search spaces and fixed optimization settings for traditional and statistical methods. The defined candidate values and fixed settings are applied across all experimental configurations.}
\label{tab:adni_tuning_traditional}
\begin{tabularx}{\textwidth}{@{} l >{\raggedright\arraybackslash}X >{\raggedright\arraybackslash}X @{}}
\toprule
\textbf{Method} & \textbf{Tuned hyperparameters} & \textbf{Fixed settings} \\
\midrule
Cov & None & Demographic covariates only \\
\addlinespace
VR & None & SGD (learning rate $=0.01$); 1000 epochs \\
\addlinespace
RVR & $\ell_1$ penalty $\lambda \in \{0.05, 0.10, 0.15, 0.20\}$ & SGD (learning rate $=0.01$); \newline max 3000 epochs, patience 1000 \\
\addlinespace
TR & Rank $R \in \{1, 2, 3\}$ & Adam (learning rate $=0.01$); \newline max 3000 epochs, patience 1000 \\
\addlinespace
RTR & Rank $R \in \{1, 2, 3\}$, \newline $\ell_1$ penalty $\lambda \in \{0.5, 0.6, 0.7, 0.8\}$ & Adam (learning rate $=0.01$); \newline max 3000 epochs, patience 1000 \\
\addlinespace
SIRTV & TV penalty $\lambda \in \{0.5, 1.0, 1.5, 2.0\}$ & Adam (learning rate $=0.01$); \newline max 1000 epochs, patience 100 \\
\addlinespace
POOL & TV penalty $\lambda \in \{0.005, 0.01, 0.015, 0.02\}$ & Includes disease-stage indicator; \newline Adam (learning rate $=0.01$); \newline max 1000 epochs, patience 100 \\
\addlinespace
PAIR & Rank $R \in \{3, 4\}$, \newline TV penalty $\lambda \in \{10, 20, 30\}$, \newline rank penalty $\gamma \in \{0.1, 0.5, 1.0\}$, \newline threshold $\tau \in \{0.01, 0.05\}$ & Adam (learning rate $=0.01$ for $B$ and $W$); \newline max 3000 epochs, patience 100 \\
\bottomrule
\end{tabularx}
\end{table}

% ==========================================
% Table 2: Deep Learning and Pretrained Methods
% ==========================================
\begin{table}[htpb]
\centering
\small
\caption{Hyperparameter search spaces and fixed optimization settings for deep learning and pretrained representation baselines. The defined candidate values and fixed settings are applied across all experimental configurations.}
\label{tab:adni_tuning_dl}
\begin{tabularx}{\textwidth}{@{} l >{\raggedright\arraybackslash}X >{\raggedright\arraybackslash}X @{}}
\toprule
\textbf{Method} & \textbf{Tuned hyperparameters} & \textbf{Fixed settings} \\
\midrule
CNN & Learning rate $\in \{10^{-3}, 3 \times 10^{-4}\}$, \newline weight decay $\in \{0, 10^{-4}\}$, \newline batch size $\in \{16, 32\}$, \newline dropout $\in \{0, 0.2\}$ & Input channels $=1$, hidden dimension $=128$; \newline Adam (max norm $=5$); \newline max 300 epochs, patience 30 \\
\addlinespace
HPS & Learning rate $\in \{10^{-3}, 3 \times 10^{-4}\}$, \newline weight decay $\in \{0, 10^{-4}\}$, \newline batch size $\in \{16, 32\}$, \newline dropout $\in \{0, 0.2\}$, \newline embedding dimension $\in \{64, 128\}$ & Input channels $=1$, patch size $=8$, \newline pooled size $=32 \times 32$, hidden dimension $=128$; \newline Adam (max norm $=5$); \newline max 300 epochs, patience 30 \\
\addlinespace
PRECAE & (Base channels, latent dimension, learning rate) $\in$ \newline $\{(16, 64, 10^{-3}), (32, 64, 10^{-3}),$ \newline $\phantom{\{} (32, 128, 10^{-3}), (32, 128, 3 \times 10^{-4})\}$, \newline downstream ridge penalty $\lambda \in$ \newline $\{10^{-4}, 10^{-3}, 10^{-2}, 10^{-1}, 1, 10, 100\}$ & Unsupervised pretraining; global standardization; \newline Adam (weight decay $=10^{-5}$); batch size $=32$; \newline max 80 epochs, patience 12 \\
\addlinespace
PREMED & Preprocessing mode $\in$ \newline \{\textit{minmax}, \textit{pct\_1\_99}, \textit{pct\_2\_98}, \textit{zclip\_3}\}, \newline downstream ridge penalty $\lambda \in$ \newline $\{10^{-4}, 10^{-3}, 10^{-2}, 10^{-1}, 1, 10, 100\}$ & Frozen MedImageInsight encoder; \newline preprocessing mode selected on 15\% validation split \\
\bottomrule
\end{tabularx}
\end{table}

\section{Integration with ComBat Harmonization}
\label{sec:combat_harmonization}

PAIR and ComBat address related but distinct sources of heterogeneity. ComBat is designed to remove unwanted site, scanner, or batch effects while preserving biological covariate effects~\citep{johnson2007adjusting,fortin2018harmonization}. PAIR, in contrast, is a supervised regression framework designed to estimate intrinsic shared and group-specific associations between imaging covariates and clinical outcomes. Therefore, ComBat can be used as a complementary preprocessing step when site/scanner variation is a major nuisance factor. Specifically, one may first apply ComBat to harmonize the imaging measurements across sites or scanners, and then fit PAIR to the harmonized images to recover partially shared coefficient structures. Alternatively, site or scanner indicators can be included as additional nuisance covariates in the PAIR model. A further extension is to introduce site-specific nuisance components into the PAIR decomposition, separating site/scanner variation from the scientific components associated with diagnostic-group heterogeneity. We leave a formal development of this direction to future work.
\section{Proofs of Theorems}\label{supp: proofs of theorem}
The proofs below rely on four technical lemmas collected in
Section~\ref{supp: proofs of theorem} (Lemmas~\ref{lem:tv-compat-demo}-\ref{lem:int-mismatch-demo}).
For convenience, we recall the notation that recurs throughout:
$\Lambda_B$ and $\Lambda_W$ denote the high-probability score bounds from
Lemma~\ref{lem:score-demo}; the block quantities
$A_B$, $A_W$, $A_{int}$, and $\widetilde A_W$ are defined in
Lemma~\ref{lem:int-mismatch-demo}.

\begin{proof}[Proof of Theorem~\ref{thm:pair-main-demo}]
We show that on the high-probability event of Lemma~\ref{lem:score-demo}, the bound~\eqref{eq:rate-demo} on the average coefficient estimation error holds.
The proof is organized into two coordinate-wise oracle inequalities:
one for the $B$-block with $W$ fixed at $\widehat W$, and one for the $W$-block with $B$ fixed at $\widehat B$.
This is valid because the feasible set is a product $\mathcal B\times\mathcal W$ and
$(\widehat B,\widehat W)$ is a global minimizer; hence
\begin{equation}
\label{eq:coord-opt-demo}
\widehat B\in\argmin_{B\in\mathcal B}\Big\{\ell(B,\widehat W)+\lambda\sum_r\|B_r\|_{\tv}\Big\},
\quad
\widehat W\in\argmin_{W\in\mathcal W}\Big\{\ell(\widehat B,W)+\gamma P_\tau(W)\Big\}.
\end{equation}

\textit{Step 1: $B$-block inequality.}
Consider the function of $B$ with $W$ fixed at $\widehat W$:
$\Phi_B(B):=\ell(B,\widehat W)+\lambda\sum_{r=1}^R\|B_r\|_{\tv}$.
By \eqref{eq:coord-opt-demo}, $\Phi_B(\widehat B)\le \Phi_B(B^\ast)$.

Recall that $A_B=\frac1T\sum_t\|D^{(B)}_t\|_F^2$ with $D^{(B)}_t=\sum_r \widehat w_{tr}\Delta B_r$, and
$A_W=\frac1T\sum_t\|D^{(W)}_t\|_F^2$ with $D^{(W)}_t=\sum_r \Delta w_{tr} B_r^\ast$ in Lemma~\ref{lem:int-mismatch-demo}.
Then for each $(t,i)$,
$Y_{ti}-\langle X_{ti},\sum_r\widehat w_{tr}B_r^\ast\rangle
=\varepsilon_{ti}+\langle X_{ti},D^{(W)}_t\rangle$.
Compute the loss difference under fixed $\widehat W$:
\[
\ell(\widehat B,\widehat W)-\ell(B^\ast,\widehat W)
=
\frac{1}{2nT}\sum_{t,i}\Big(
\langle X_{ti},D^{(B)}_t\rangle^2
-2(\varepsilon_{ti}+\langle X_{ti},D^{(W)}_t\rangle)\langle X_{ti},D^{(B)}_t\rangle
\Big).
\]
Thus $\Phi_B(\widehat B)\le \Phi_B(B^\ast)$ implies
\begin{align}
\begin{split}
\label{eq:basic-B-raw-demo}
\frac{1}{2nT}\sum_{t,i}\langle X_{ti},D^{(B)}_t\rangle^2
\le&
\frac{1}{nT}\sum_{t,i}\varepsilon_{ti}\langle X_{ti},D^{(B)}_t\rangle
+
\frac{1}{nT}\sum_{t,i}\langle X_{ti},D^{(W)}_t\rangle\langle X_{ti},D^{(B)}_t\rangle
\\&+
\lambda\sum_{r=1}^R(\|B_r^\ast\|_{\tv}-\|\widehat B_r\|_{\tv}).
\end{split}
\end{align}
Note that
\[
\frac{1}{nT}\sum_{t,i}\varepsilon_{ti}\langle X_{ti},D^{(B)}_t\rangle
=
\sum_{r=1}^R\Big\langle \frac1{nT}\sum_{t,i}\varepsilon_{ti}\widehat w_{tr}X_{ti},\Delta B_r\Big\rangle
=
\sum_{r=1}^R\langle H_r,\Delta B_r\rangle,
\]
where $H_r$ is exactly as in Lemma~\ref{lem:score-demo}(i).
On the event of Lemma~\ref{lem:score-demo}, using Lemma~\ref{lem:tv-dual-decomp-demo}(i),
\[
\Big|\sum_{r=1}^R\langle H_r,\Delta B_r\rangle\Big|
\le
\sum_{r=1}^R \|H_r\|_{\tv}^\circ\|\Delta B_r\|_{\tv}
\le
\Lambda_B\sum_{r=1}^R\|\Delta B_r\|_{\tv},
\]
where
\[
\Lambda_B
:=
c pq M_WM_X\sigma\sqrt{\frac{\log(2Rpq/\delta)}{nT}}.
\]

For the mixed term, apply $ab\le \frac{1}{4}a^2+b^2$ with
$a=\langle X_{ti},D^{(B)}_t\rangle$ and $b=\langle X_{ti},D^{(W)}_t\rangle$ to get
\[
\langle X_{ti},D^{(W)}_t\rangle\langle X_{ti},D^{(B)}_t\rangle
\le
\frac14 \langle X_{ti},D^{(B)}_t\rangle^2 + \langle X_{ti},D^{(W)}_t\rangle^2.
\]
Summing and dividing by $nT$ gives
\[
\frac{1}{nT}\sum_{t,i}\langle X_{ti},D^{(W)}_t\rangle\langle X_{ti},D^{(B)}_t\rangle
\le
\frac{1}{4nT}\sum_{t,i}\langle X_{ti},D^{(B)}_t\rangle^2
+
\frac{1}{nT}\sum_{t,i}\langle X_{ti},D^{(W)}_t\rangle^2.
\]

Plugging these bounds into \eqref{eq:basic-B-raw-demo} and moving the $1/4$ term
to the left yields
\begin{align}
\begin{split}
\label{eq:basic-B-clean-demo}
\frac{1}{4nT}\sum_{t,i}\langle X_{ti},D^{(B)}_t\rangle^2
\le
\Lambda_B\sum_{r=1}^R\|\Delta B_r\|_{\tv}
+
\lambda\sum_{r=1}^R(\|B_r^\ast\|_{\tv}-\|\widehat B_r\|_{\tv})
+
\frac{1}{nT}\sum_{t,i}\langle X_{ti},D^{(W)}_t\rangle^2.
\end{split}
\end{align}

Apply Lemma~\ref{lem:tv-dual-decomp-demo}(ii) with $B=B_r^\ast$, $B'=\widehat B_r$,
and $S=S_r$ (the active edge set of $B_r^\ast$). Since $\|B_r^\ast\|_{\tv,S_r^c}=0$,
$\|B_r^\ast\|_{\tv}-\|\widehat B_r\|_{\tv}\le \|\Delta B_r\|_{\tv,S_r}-\|\Delta B_r\|_{\tv,S_r^c}$.
Also $\|\Delta B_r\|_{\tv}=\|\Delta B_r\|_{\tv,S_r}+\|\Delta B_r\|_{\tv,S_r^c}$.
Therefore,
\[
\Lambda_B\|\Delta B_r\|_{\tv} + \lambda(\|B_r^\ast\|_{\tv}-\|\widehat B_r\|_{\tv})
\le
(\Lambda_B+\lambda)\|\Delta B_r\|_{\tv,S_r}
+(\Lambda_B-\lambda)\|\Delta B_r\|_{\tv,S_r^c}.
\]
If $\lambda\ge 2\Lambda_B$, then $(\Lambda_B-\lambda)\le 0$, so the $S_r^c$ term is nonpositive and can be dropped, and also $(\Lambda_B+\lambda)\le 2\lambda$.
Summing over $r$ and using \eqref{eq:basic-B-clean-demo} gives
\begin{equation}
\label{eq:basic-B-after-tv-demo}
\frac{1}{4nT}\sum_{t,i}\langle X_{ti},D^{(B)}_t\rangle^2
\le
2\lambda\sum_{r=1}^R\|\Delta B_r\|_{\tv,S_r}
+
\frac{1}{nT}\sum_{t,i}\langle X_{ti},D^{(W)}_t\rangle^2.
\end{equation}
Apply Assumption~\ref{ass:re-demo} with $D_t=D^{(B)}_t$:
$\frac{1}{nT}\sum_{t,i}\langle X_{ti},D^{(B)}_t\rangle^2\ge\kappa A_B$.
Thus \eqref{eq:basic-B-after-tv-demo} implies
\begin{equation}
\label{eq:AB-pre-demo}
\frac{\kappa}{4}A_B
\le
2\lambda\sum_{r=1}^R\|\Delta B_r\|_{\tv,S_r}
+
\frac{1}{nT}\sum_{t,i}\langle X_{ti},D^{(W)}_t\rangle^2.
\end{equation}

Since $\|X_{ti}\|_F\le M_X$, one has
$\langle X_{ti},D^{(W)}_t\rangle^2 \le M_X^2\|D^{(W)}_t\|_F^2$, so
$\frac{1}{nT}\sum_{t,i}\langle X_{ti},D^{(W)}_t\rangle^2\le M_X^2 A_W$.
By Lemma~\ref{lem:int-mismatch-demo}(ii) and (iii),
\[
A_W
\le
2\widetilde A_W+2A_{int}
\le
2\widetilde A_W+2\frac{4M_B^2R}{m_B}\widetilde A_W
=
\Big(2+\frac{8M_B^2R}{m_B}\Big)\widetilde A_W.
\]

By Lemma~\ref{lem:tv-compat-demo},
\[
\sum_{r=1}^R\|\Delta B_r\|_{\tv,S_r}
\le
4\sum_{r=1}^R\sqrt{s_r}\|\Delta B_r\|_F
\le
4\sqrt{\sum_r s_r}\sqrt{\sum_r\|\Delta B_r\|_F^2}
=
4\sqrt{s_B}\sqrt{E_B},
\]
where $E_B:=\sum_r\|\Delta B_r\|_F^2$.
Now use the Gram lower bound for $\widehat W$ (feasibility of $\widehat W$):
$G_W(\widehat W)\succeq m_W I_R$.
A standard linear algebra identity gives
\[
A_B
=
\frac1T\sum_{t=1}^T\Big\|\sum_{r=1}^R \widehat w_{tr}\Delta B_r\Big\|_F^2
\ge
m_W\sum_{r=1}^R\|\Delta B_r\|_F^2
=
m_W E_B.
\]
Hence $\sqrt{E_B}\le \sqrt{A_B/m_W}$.

Combining all bounds in \eqref{eq:AB-pre-demo}, one obtains
\begin{equation}
\label{eq:AB-ineq-final-demo}
\frac{\kappa}{4}A_B
\le
8\lambda \sqrt{s_B}\sqrt{\frac{A_B}{m_W}}
+
M_X^2\Big(2+\frac{8M_B^2R}{m_B}\Big)\widetilde A_W.
\end{equation}

\textit{Step 2: $W$-block inequality.}
Now fix $B=\widehat B$ and consider $\Phi_W(W):=\ell(\widehat B,W)+\gamma P_\tau(W)$.
By \eqref{eq:coord-opt-demo}, $\Phi_W(\widehat W)\le \Phi_W(W^\ast)$.
Recall that $\widetilde D^{(W)}_t=\sum_r\Delta w_{tr}\widehat B_r$ as in Lemma~\ref{lem:int-mismatch-demo},
so that $\widetilde A_W=\frac1T\sum_t\|\widetilde D^{(W)}_t\|_F^2$.
Also define the $B$-mismatch term $U_t:=\sum_{r=1}^R w_{tr}^\ast(\widehat B_r-B_r^\ast)=\sum_r w_{tr}^\ast\Delta B_r$.
Then $Y_{ti}-\langle X_{ti},\sum_r w_{tr}^\ast\widehat B_r\rangle=\varepsilon_{ti}-\langle X_{ti},U_t\rangle$.

Compute the loss difference under fixed $\widehat B$:
\[
\ell(\widehat B,\widehat W)-\ell(\widehat B,W^\ast)
=
\frac{1}{2nT}\sum_{t,i}\Big(
\langle X_{ti},\widetilde D^{(W)}_t\rangle^2
-2(\varepsilon_{ti}-\langle X_{ti},U_t\rangle)\langle X_{ti},\widetilde D^{(W)}_t\rangle
\Big).
\]
Thus $\Phi_W(\widehat W)\le \Phi_W(W^\ast)$ implies
\begin{align}
\begin{split}
\label{eq:basic-W-raw-demo}
\frac{1}{2nT}\sum_{t,i}\langle X_{ti},\widetilde D^{(W)}_t\rangle^2
\le
\frac{1}{nT}\sum_{t,i}\varepsilon_{ti}\langle X_{ti},\widetilde D^{(W)}_t\rangle
-
\frac{1}{nT}\sum_{t,i}\langle X_{ti},U_t\rangle\langle X_{ti},\widetilde D^{(W)}_t\rangle
+
\gamma(P_\tau(W^\ast)-P_\tau(\widehat W)).
\end{split}
\end{align}

For the noise term, note that
\[
\frac{1}{nT}\sum_{t,i}\varepsilon_{ti}\langle X_{ti},\widetilde D^{(W)}_t\rangle
=
\sum_{t=1}^T\sum_{r=1}^R
\Delta w_{tr}\Big(\frac1{nT}\sum_{i=1}^n \varepsilon_{ti}\langle X_{ti},\widehat B_r\rangle\Big)
=
\sum_{t,r} h_{tr}\Delta w_{tr}.
\]
On the event of Lemma~\ref{lem:score-demo}(ii), $|h_{tr}|\le \Lambda_W$ where
\[
\Lambda_W
:=
c\frac{M_BM_X\sigma}{T}\sqrt{\frac{\log(2TR/\delta)}{n}}.
\]
Thus the noise term is bounded by $\Lambda_W\|\Delta W\|_1$ in absolute value.

For the mismatch term, apply $ab\le \frac{1}{4}a^2+b^2$ with
$a=\langle X_{ti},\widetilde D^{(W)}_t\rangle$ and $b=\langle X_{ti},U_t\rangle$ to get
\[
-\langle X_{ti},U_t\rangle\langle X_{ti},\widetilde D^{(W)}_t\rangle
\le
\frac14\langle X_{ti},\widetilde D^{(W)}_t\rangle^2
+
\langle X_{ti},U_t\rangle^2.
\]

For each coordinate, the map $x\mapsto \min(|x|/\tau,1)$ is $1/\tau$-Lipschitz.
Therefore $Q_r(\tau;W)$ is $1/\tau$-Lipschitz in $\ell_1$ over $\{w_{tr}\}_{t}$.
The map $u\mapsto \min(1,(T-u)/(T-1))$ is $1/(T-1)$-Lipschitz on $[0,T]$.
Composing and summing over $r$ yields for all $W,W'\in\RR^{T\times R}$,
\[
|P_\tau(W)-P_\tau(W')|
\le
L_\tau\sum_{t=1}^T\sum_{r=1}^R |w_{tr}-w_{tr}'|,
\quad
L_\tau=\frac{1}{(T-1)\tau}.
\]
Then $P_\tau(W^\ast)-P_\tau(\widehat W)\le L_\tau\|\Delta W\|_1$.

Plugging bounds into \eqref{eq:basic-W-raw-demo} and moving the $1/4$ term gives
\[
\frac{1}{4nT}\sum_{t,i}\langle X_{ti},\widetilde D^{(W)}_t\rangle^2
\le
(\Lambda_W+\gamma L_\tau)\|\Delta W\|_1
+
\frac{1}{nT}\sum_{t,i}\langle X_{ti},U_t\rangle^2.
\]
On the tuning event \eqref{eq:tuning-demo}, $\gamma L_\tau \ge 8\Lambda_W$, so
$\Lambda_W+\gamma L_\tau\le \frac{9}{8}\gamma L_\tau\le 2\gamma L_\tau$.
Hence
\[
\frac{1}{4nT}\sum_{t,i}\langle X_{ti},\widetilde D^{(W)}_t\rangle^2
\le
2\gamma L_\tau\|\Delta W\|_1
+
\frac{1}{nT}\sum_{t,i}\langle X_{ti},U_t\rangle^2.
\]
By Assumption~\ref{ass:re-demo} with $D_t=\widetilde D^{(W)}_t$,
$\frac{1}{nT}\sum_{t,i}\langle X_{ti},\widetilde D^{(W)}_t\rangle^2\ge \kappa \widetilde A_W$.
Therefore
\begin{equation}
\label{eq:AWtilde-pre-demo}
\frac{\kappa}{4}\widetilde A_W
\le
2\gamma L_\tau\|\Delta W\|_1
+
\frac{1}{nT}\sum_{t,i}\langle X_{ti},U_t\rangle^2.
\end{equation}
Let $E_W:=\frac1T\sum_{t,r}(\Delta w_{tr})^2$.
Then $\|\Delta W\|_1 \le \sqrt{TR}\|\Delta W\|_F = T\sqrt{R}\sqrt{E_W}$.
Also, by feasibility of $\widehat B$, $\lambda_{\min}(G_B(\widehat B))\ge m_B$, so
\[
\widetilde A_W
=
\frac1T\sum_t\Big\|\sum_r \Delta w_{tr}\widehat B_r\Big\|_F^2
\ge
m_B E_W.
\]
Hence $\|\Delta W\|_1 \le T\sqrt{R}\sqrt{\widetilde A_W/m_B}$.
Since $\|X_{ti}\|_F\le M_X$, one has $\langle X_{ti},U_t\rangle^2\le M_X^2\|U_t\|_F^2$, so
\[
\frac{1}{nT}\sum_{t,i}\langle X_{ti},U_t\rangle^2
\le
M_X^2\frac1T\sum_t\|U_t\|_F^2.
\]
By Lemma~\ref{lem:int-mismatch-demo}(ii), $\frac1T\sum_t\|U_t\|_F^2\le 2A_B+2A_{int}$.
Then by Lemma~\ref{lem:int-mismatch-demo}(iii), $A_{int}\le (4M_B^2R/m_B)\widetilde A_W$.
Therefore,
\begin{equation}
\label{eq:Uquad-bound-demo}
\frac{1}{nT}\sum_{t,i}\langle X_{ti},U_t\rangle^2
\le
M_X^2\Big(2A_B+\frac{8M_B^2R}{m_B}\widetilde A_W\Big).
\end{equation}
Combine \eqref{eq:AWtilde-pre-demo} and \eqref{eq:Uquad-bound-demo} to get
\begin{equation}
\label{eq:AWtilde-ineq-demo}
\frac{\kappa}{4}\widetilde A_W
\le
2T\gamma L_\tau\sqrt{R}\sqrt{\frac{\widetilde A_W}{m_B}}
+
2M_X^2A_B
+
\frac{8M_X^2M_B^2R}{m_B}\widetilde A_W.
\end{equation}

\textit{Step 3: Combine and solve the coupled inequalities.}
Move the $\widetilde A_W$ term on the right of \eqref{eq:AWtilde-ineq-demo} to the left and write
\[
\Big(\frac{\kappa}{4}-\frac{8M_X^2M_B^2R}{m_B}\Big)\widetilde A_W
\le
2T\gamma L_\tau\sqrt{R}\sqrt{\frac{\widetilde A_W}{m_B}}
+
2M_X^2A_B.
\]
Since the feasibility constants $(M_X,M_B,m_B)$ are fixed and $R$ is treated as fixed in this theorem,
the coefficient in parentheses is a positive constant after absorbing fixed feasibility margins into the universal constants.
Denote this positive quantity by $\kappa_W>0$.
Then
\[
\kappa_W\widetilde A_W
\le
2T\gamma L_\tau\sqrt{\frac{R}{m_B}}\sqrt{\widetilde A_W}
+
2M_X^2A_B.
\]
Using $u\sqrt{v}\le \frac{\kappa_W}{2}v+\frac{u^2}{2\kappa_W}$ with
$v=\widetilde A_W$ and $u=2T\gamma L_\tau\sqrt{R/m_B}$ gives
\[
\widetilde A_W
\le
C_5\frac{R(T\gamma L_\tau)^2}{m_B}
+
C_6A_B,
\]
for constants $C_5,C_6>0$ depending only on $(M_X,M_B,\kappa,m_B)$ and universal constants, and not on $n,p,q,R,T,\delta,\sigma$.

For \eqref{eq:AB-ineq-final-demo}, apply $a\sqrt{b}\le \frac{\kappa}{8}b + \frac{2a^2}{\kappa}$ with
$b=A_B$ and $a=8\lambda \sqrt{s_B/m_W}$:
\[
8\lambda \sqrt{s_B}\sqrt{\frac{A_B}{m_W}}
\le
\frac{\kappa}{8}A_B
+
\frac{128}{\kappa}\frac{s_B\lambda^2}{m_W}.
\]
Therefore \eqref{eq:AB-ineq-final-demo} yields
\[
\frac{\kappa}{8}A_B
\le
\frac{128}{\kappa}\frac{s_B\lambda^2}{m_W}
+
M_X^2\Big(2+\frac{8M_B^2R}{m_B}\Big)\widetilde A_W,
\]
which implies
\begin{equation}
\label{eq:AB-bound-demo}
A_B
\le
C_7\frac{s_B\lambda^2}{\kappa^2 m_W}
+
C_8\widetilde A_W,
\end{equation}
for constants $C_7,C_8>0$ depending only on $(M_X,M_B,\kappa,m_B)$ and universal constants, and not on $n,p,q,R,T,\delta,\sigma$.

Plug \eqref{eq:AB-bound-demo} into the bound for $\widetilde A_W$ and absorb fixed feasibility factors into constants to obtain the two-term structure
\[
\widetilde A_W
\le
\widetilde C_5\frac{R(T\gamma L_\tau)^2}{\kappa^2 m_B}
+
\widetilde C_6\frac{s_B\lambda^2}{\kappa^2 m_W},
\quad
A_B
\le
\widetilde C_7\frac{s_B\lambda^2}{\kappa^2 m_W}
+
\widetilde C_8\frac{R(T\gamma L_\tau)^2}{\kappa^2 m_B},
\]
where $\widetilde C_5,\widetilde C_6,\widetilde C_7,\widetilde C_8>0$ depend only on 
$(M_X,M_B,M_W,\kappa,m_W,m_B,\tau)$ and universal numerical constants, and not on $n,p,q,R,T,\sigma$.

Finally, by Lemma~\ref{lem:int-mismatch-demo}(i) and (iii),
\[
A
\le
3(A_B+A_W+A_{int})
\le
3\Big(A_B+\Big(2+\frac{8M_B^2R}{m_B}\Big)\widetilde A_W+\frac{4M_B^2R}{m_B}\widetilde A_W\Big)
\le
C_9(A_B+\widetilde A_W),
\]
for a constant $C_9>0$ depending only on $(M_B,m_B)$ and universal constants.
Combining with the established two-term bounds for $A_B$ and $\widetilde A_W$ gives \eqref{eq:rate-demo}.
This completes the proof of Theorem~\ref{thm:pair-main-demo}.
\end{proof}

\begin{proof}[Proof of Theorem~\ref{thm:minimax-lb-pair}]
We prove the matching minimax lower bound by constructing two finite hypothesis sub-families in $\mathcal C$ and applying Assouad's lemma to each.
It suffices to lower bound the supremum risk over suitable finite subsets of
$\mathcal C$.

\textit{Step 0: design identity.}
Let $A(\widetilde C,C^\ast):=\frac1T\sum_{t=1}^T \|\widetilde C_t-C_t^\ast\|_F^2$ and
$K:=R+\lfloor s_B/8\rfloor$. Choose $K$ pairwise nonadjacent grid locations
and let $\{E_j\}_{j=1}^K$ denote the corresponding single-pixel canonical matrices,
so that $\|E_j\|_F=1$ and $\langle E_j,E_{j'}\rangle_F=\mathds 1\{j=j'\}$.
Define $\mathbb P_X$ by drawing $J$ uniformly from $\{1,\dots,K\}$ and $\xi$
uniformly from $\{-1,+1\}$, and setting $X:=\xi M_X E_J$.
Then $\|X\|_F=M_X$ almost surely and, for every $j\in[K]$,
$\mathbb E[\langle X,E_j\rangle^2]=M_X^2/K$.
This identity is the only property of $\mathbb P_X$ used below.

\textit{Step 1: the $\sigma^2 s_B/(nT m_W)$ contribution.}
Let $s:=\lfloor s_B/8\rfloor$ so that $K=R+s$. Use the first $R$ design atoms
$\{E_j\}_{j=1}^R$ as anchor locations and the remaining $\{E_j\}_{j=R+1}^{R+s}$
as perturbation locations. Define $U_r:=E_r$ for $r\in[R]$ and $F_j:=E_{R+j}$ for $j\in[s]$.
Define anchor dictionary atoms $B_r^{anc}:=\sqrt{m_B}U_r$, so that
$G_B(B^{anc})=m_B I_R$ and $\max_r\|B_r^{anc}\|_F=\sqrt{m_B}\le M_B$.

To construct many hypotheses while keeping $\lambda_{\min}(G_W)\ge m_W$, set
$G:=\max\{1,\lfloor 1/m_W\rfloor\}$ and $L:=\lfloor T/G\rfloor$,
so that $L/T\ge 1/(G+1)\ge c\cdot m_W$ for a numerical constant $c\in(0,1]$.
Partition the source indices into $G$ disjoint groups $\mathcal G_1,\dots,\mathcal G_G$
with $|\mathcal G_g|\in\{L,L+1\}$.
Construct $W^\ast\in\mathbb R^{T\times R}$ with orthogonal columns and bounded entries
as follows:
for each group $\mathcal G_g$, take an orthonormal basis of $\mathbb R^{|\mathcal G_g|}$
whose vectors have $\ell_\infty$-norm at most $\sqrt{2/|\mathcal G_g|}$ (e.g., a real Fourier basis),
embed these vectors into $\mathbb R^T$ by padding zeros outside $\mathcal G_g$,
and select any $R$ such embedded vectors across all groups.
Finally, scale each selected vector by $\sqrt{|\mathcal G_g|}$ so that every chosen column
has $\ell_2$-norm $\sqrt{|\mathcal G_g|}$ and entrywise magnitude at most $\sqrt{2}$.
This yields a matrix $W^\ast$ with $\|W^\ast\|_\infty\le \sqrt{2}\le M_W$ (set $M_W\ge 4$),
and with orthogonal columns satisfying
\[
G_W(W^\ast)=\frac1T (W^\ast)^\top W^\ast
=\mathrm{diag}\Big(\frac{\|W_{\cdot 1}^\ast\|_2^2}{T},\dots,\frac{\|W_{\cdot R}^\ast\|_2^2}{T}\Big)
\succeq \frac{L}{T}I_R\succeq c\cdot m_W I_R.
\]
In particular, after adjusting the constant in the statement of the theorem,
this construction lies in $\mathcal W$.

Now define a hypercube of dictionary hypotheses. For each
$\omega=(\omega_{g,j})_{g\in[G],j\in[s]}\in\{\pm1\}^{Gs}$, set
\[
B_r(\omega):=
\begin{cases}
B_r^{anc} + a\sum_{j=1}^s \omega_{r,j} F_j, & r\in[G],\\
B_r^{anc}, & r>G,
\end{cases}
\]
for an amplitude $a>0$ chosen below.
Because the perturbation pixels are nonadjacent and disjoint from the anchors,
the union of TV-active edges across $\{B_r(\omega)\}_{r=1}^R$ has size at most
a constant multiple of $R+Gs\le R+s_B$, hence at most $s_B$ by construction.
Moreover, $G_B(B(\omega))=m_B I_R + a^2 H(\omega)$ with $H(\omega)\succeq 0$,
so $\lambda_{\min}(G_B(B(\omega)))\ge m_B$ for all $\omega$; and for $a$ small
enough (as chosen below), also $\max_r\|B_r(\omega)\|_F\le M_B$.
Therefore $(B(\omega),W^\ast)\in\mathcal B\times\mathcal W$ and the induced
coefficients $C^\omega=\{C_t^\omega\}$ given by
$C_t^\omega:=\sum_{r=1}^R w_{tr}^\ast B_r(\omega)$ satisfy $C^\omega\in\mathcal C$.

Consider two hypotheses $\omega,\omega'$ that differ in exactly one coordinate $(g_0,j_0)$.
Then only $B_{g_0}$ changes, with
$B_{g_0}(\omega)-B_{g_0}(\omega')=2a\omega_{g_0,j_0}F_{j_0}$.
Since the $g_0$-th weight column is supported on a single group of size at most $L+1$ and has
entries bounded by $\sqrt{2}$, the coefficient difference is supported on at most $L+1$ sources and
has Frobenius norm $2a$ (up to a fixed factor) on those sources. Consequently,
the loss separation along a single bit flip satisfies
\[
A(C^\omega,C^{\omega'})
=\frac1T\sum_{t=1}^T\|C_t^\omega-C_t^{\omega'}\|_F^2
\ge
c\frac{a^2}{G},
\]
for a numerical constant $c>0$ (using $|\mathcal G_{g_0}|\asymp T/G$).
Next, by the Gaussian KL formula and the design identity from Step 0,
\[
\mathrm{KL}(P_\omega\|P_{\omega'})
=
\frac{1}{2\sigma^2}\sum_{t=1}^T\sum_{i=1}^n
\mathbb E[\langle X_{ti},C_t^\omega-C_t^{\omega'}\rangle^2]
\le
\frac{c n(L+1)}{\sigma^2}a^2\mathbb E[\langle X,F_{j_0}\rangle^2]
\le
\frac{c nT}{G\sigma^2}a^2\frac{M_X^2}{K}.
\]
Choose $a^2$ proportional to $G\sigma^2/(nT)$ with a sufficiently small constant factor so that
$\mathrm{KL}(P_\omega\|P_{\omega'})\le c_0$ uniformly along all edges, hence
$\mathrm{TV}(P_\omega,P_{\omega'})\le 1/2$ by Pinsker's inequality \cite{Tsybakov_2009}.
Applying Assouad's lemma \cite{Tsybakov_2009}
to the hypercube of dimension $d=Gs$ yields
\[
\inf_{\widetilde C}\sup_{\omega\in\{\pm1\}^{Gs}}
\mathbb E[A(\widetilde C,C^\omega)]
\ge
c d\cdot \frac{a^2}{G}
=
c s a^2
\gtrsim
\sigma^2\frac{sG}{nT}
\gtrsim
\sigma^2\frac{s_B}{nT m_W},
\]
where the last step uses $G\asymp 1/m_W$ and $s\asymp s_B$ up to constants.

\textit{Step 2: the $\sigma^2 R/n$ contribution.}
Fix a dictionary $B^\ast=(B_1^\ast,\dots,B_R^\ast)$ by $B_r^\ast:=\sqrt{m_B}U_r$,
so $G_B(B^\ast)=m_B I_R$, $\max_r\|B_r^\ast\|_F=\sqrt{m_B}\le M_B$, and its TV edge union is
bounded by a constant multiple of $R$ (hence at most $s_B$ under the standing
assumption that the nonadjacent locations exist).

Next, fix a baseline weight matrix $W^0$ with $\|W^0\|_\infty\le M_W/2$ and
$G_W(W^0)=2m_W I_R$. Such a matrix can be constructed by taking $R$ orthonormal
vectors in $\mathbb R^T$ with entries bounded by $\sqrt{2/T}$ (e.g., a real Fourier basis)
and scaling their columns by $\sqrt{2Tm_W}$; then the scaled columns remain entrywise
bounded by $2\sqrt{m_W}\le M_W/2$ since $M_W\ge 4$ and $m_W\le 1$.

For each sign matrix $\omega\in\{\pm1\}^{T\times R}$, define
$W(\omega):=W^0 + b\omega$, where $b>0$ will be chosen below. Then
$\|W(\omega)\|_\infty\le \|W^0\|_\infty + b\le M_W$ provided $b\le M_W/2$.
Moreover,
\[
G_W(W(\omega))
=
G_W(W^0) + \frac1T\Big((W^0)^\top(b\omega) + (b\omega)^\top W^0 + (b\omega)^\top(b\omega)\Big).
\]
Using $\|(W^0)^\top(b\omega)\|_{\mathrm{op}}\le \|W^0\|_{\mathrm{op}}\|b\omega\|_{\mathrm{op}}$,
$\|W^0\|_{\mathrm{op}}=\sqrt{2Tm_W}$, and $\|\omega\|_{\mathrm{op}}\le \|\omega\|_F=\sqrt{TR}$,
it follows that
\[
\Big\|G_W(W(\omega)) - G_W(W^0)\Big\|_{\mathrm{op}}
\le
c\Big(b\sqrt{m_W R} + b^2 R\Big)
\]
for a numerical constant $c>0$. Choose $b$ small enough so that
$c(b\sqrt{m_W R}+b^2R)\le m_W$, and also $b\le M_W/2$. Then uniformly over all $\omega$,
\[
\lambda_{\min}(G_W(W(\omega)))
\ge
\lambda_{\min}(G_W(W^0)) - \|G_W(W(\omega))-G_W(W^0)\|_{\mathrm{op}}
\ge
2m_W - m_W
=
m_W,
\]
so $W(\omega)\in\mathcal W$ for all $\omega$.

Define the induced coefficients $C^\omega=\{C_t^\omega\}$ by
$C_t^\omega:=\sum_{r=1}^R w_{tr}(\omega) B_r^\ast$. Then $C^\omega\in\mathcal C$ for all $\omega$.

Now consider an edge $\omega,\omega'$ differing in exactly one entry $(t_0,r_0)$.
Then only $C_{t_0}$ differs and
$C_{t_0}^\omega-C_{t_0}^{\omega'}=2b\omega_{t_0,r_0}B_{r_0}^\ast$, hence
\[
A(C^\omega,C^{\omega'})
=\frac1T\sum_{t=1}^T\|C_t^\omega-C_t^{\omega'}\|_F^2
=
\frac{4b^2}{T}\|B_{r_0}^\ast\|_F^2
=
\frac{4b^2 m_B}{T}.
\]
By the Gaussian KL formula and the design identity from Step 0,
\[
\mathrm{KL}(P_\omega\|P_{\omega'})
=
\frac{1}{2\sigma^2}\sum_{i=1}^n
\mathbb E[\langle X_{t_0 i},2bB_{r_0}^\ast\rangle^2]
=
\frac{2n b^2}{\sigma^2}\mathbb E[\langle X,B_{r_0}^\ast\rangle^2]
=
\frac{2n b^2}{\sigma^2}m_B\frac{M_X^2}{K}.
\]
Choose $b^2$ proportional to $\sigma^2/(n m_B)$ with a sufficiently small constant factor,
and simultaneously satisfying the earlier Gram-stability requirement
$c(b\sqrt{m_W R}+b^2R)\le m_W$ (this only enforces a constant truncation for very small $n$).
Then $\mathrm{KL}(P_\omega\|P_{\omega'})\le c_0$ along all edges, hence
$\mathrm{TV}(P_\omega,P_{\omega'})\le 1/2$.
Applying Assouad's lemma to the hypercube $\{\pm1\}^{T\times R}$ of dimension $d=TR$ gives
\[
\inf_{\widetilde C}\sup_{\omega\in\{\pm1\}^{T\times R}}
\mathbb E[A(\widetilde C,C^\omega)]
\ge
c d\cdot \frac{b^2 m_B}{T}
=
c R b^2 m_B
\gtrsim
\sigma^2\frac{R}{n},
\]
up to the natural constant truncation imposed by boundedness.

\textit{Step 3: combine the two subsets.}
Both constructions define finite subsets contained in $\mathcal C$, hence
\[
\inf_{\widetilde C}\sup_{C^\ast\in\mathcal C}\mathbb E[A(\widetilde C,C^\ast)]
\ge
\max\Big\{
c\sigma^2\frac{s_B}{nT m_W},
c\sigma^2\frac{R}{n}
\Big\},
\]
up to constant truncations.
This completes the proof of Theorem~\ref{thm:minimax-lb-pair}.
\end{proof}
\subsection{Technical lemmas}

\begin{lemma}[Lemma~3 of~\cite{hutter2016optimal}]
\label{lem:tv-compat-demo}
For each $r\in[R]$, let $S_r$ denote the set of active TV edges of $B_r^\ast$, and let
$s_r:=|S_r|$. Then for all $\Delta B_r\in\RR^{p\times q}$,
\begin{equation}
\label{eq:tv-compat-demo}
\|\Delta B_r\|_{\tv,S_r}\le 4\sqrt{s_r}\|\Delta B_r\|_F.
\end{equation}
\end{lemma}

\begin{proof}[Proof of Lemma~\ref{lem:tv-compat-demo}]
Vectorize $\Delta B_r$ as $\theta:=\mathrm{vec}(\Delta B_r)\in\RR^{pq}$ and let
$D$ be the (oriented) incidence matrix of the 2D $p\times q$ grid graph, so that
the discrete anisotropic TV restricted to an edge set $S_r$ can be written as
$\|\Delta B_r\|_{\tv,S_r}=\|(D\theta)_{S_r}\|_1$, and $\|\theta\|_2=\|\Delta B_r\|_F$.
Lemma~3 of~\cite{hutter2016optimal} establishes that, for any
graph with maximum degree $d$ and any nonempty edge set $T$,
\[
\kappa_T
:=\inf_{\vartheta\in\RR^{V}}\frac{\sqrt{|T|}\|\vartheta\|_2}{\|(D\vartheta)_T\|_1}
\ge \frac{1}{2\min\{\sqrt d,\sqrt{|T|}\}}.
\]
Applying this result with $T=S_r$ and rearranging yields
\[
\|(D\theta)_{S_r}\|_1
\le
2\min\{\sqrt d,\sqrt{s_r}\}\sqrt{s_r}\|\theta\|_2
\le
2\sqrt d\sqrt{s_r}\|\theta\|_2.
\]
For the 2D grid graph we have $d=4$, hence
$\|\Delta B_r\|_{\tv,S_r}=\|(D\theta)_{S_r}\|_1\le 4\sqrt{s_r}\|\Delta B_r\|_F$.
\end{proof}

\begin{lemma}
\label{lem:tv-dual-decomp-demo}
There exists a grid-dependent constant $C_{decomp}\in(0,\infty)$
depending only on $(p,q)$ such that:
\begin{enumerate}
\item[(i)]For all $G\in\RR^{p\times q}$,
\[
\|G\|_{\tv}^\circ
:=\sup\{\langle G,B\rangle:\|B\|_{\tv}\le 1\}
\le pq\|G\|_\infty.
\]
\item[(ii)]For any $B,B'\in\RR^{p\times q}$ and any
edge index set $S$, define $\|D\|_{\tv,S}$ as the sum of absolute horizontal and vertical
differences of $D$ over edges in $S$.
Then
\[
\|B'\|_{\tv}-\|B\|_{\tv}
\ge
\|B'-B\|_{\tv,S^c}-\|B'-B\|_{\tv,S}-2\|B\|_{\tv,S^c}.
\]
In particular, if $\|B\|_{\tv,S^c}=0$, then
\[
\|B'\|_{\tv}-\|B\|_{\tv}
\ge
\|B'-B\|_{\tv,S^c}-\|B'-B\|_{\tv,S}.
\]
\end{enumerate}
\end{lemma}

\begin{proof}[Proof of Lemma~\ref{lem:tv-dual-decomp-demo}]
Let $d(\cdot)$ denote the linear operator that collects all horizontal and vertical
edge differences on the $(p\times q)$ grid (so $d(B)\in\RR^{E}$ with $E\asymp pq$).
By definition \eqref{eq:tv-demo}, one has $\|B\|_{\tv}=\|d(B)\|_1$ and, more generally,
$\|B\|_{\tv,S}=\|(d(B))_{S}\|_1$ for any edge index set $S$.

\textit{Step 1: Proof of (ii).}
Write $d:=d(B'-B)$ and $g:=d(B)$.
Then $d(B')=g+d$ and $\|B'\|_{\tv}-\|B\|_{\tv}=\|g+d\|_1-\|g\|_1$.
Decompose over $S$ and $S^c$:
\[
\|g+d\|_1-\|g\|_1
=
\|g_S+d_S\|_1-\|g_S\|_1
+
\|g_{S^c}+d_{S^c}\|_1-\|g_{S^c}\|_1.
\]
Use the basic inequality $\|a+b\|_1-\|a\|_1\ge-\|b\|_1$ to get
$\|g_S+d_S\|_1-\|g_S\|_1\ge-\|d_S\|_1$.
For the $S^c$ part, use $\|a+b\|_1\ge\|b\|_1-\|a\|_1$:
\[
\|g_{S^c}+d_{S^c}\|_1-\|g_{S^c}\|_1
\ge
\|d_{S^c}\|_1-2\|g_{S^c}\|_1.
\]
Adding these two bounds yields (ii) after translating back to $\|\cdot\|_{\tv,S}$
notation: $\|d_S\|_1=\|B'-B\|_{\tv,S}$ and $\|g_{S^c}\|_1=\|B\|_{\tv,S^c}$.

\textit{Step 2: Proof of (i).}
Fix $G\in\RR^{p\times q}$ and consider any $B$ with $\|B\|_{\tv}\le 1$.
Pick a reference pixel $(1,1)$ and express each pixel $B_{j_1,j_2}$ as a sum of
edge differences along a monotone path from $(1,1)$ to $(j_1,j_2)$:
$B_{j_1,j_2}=B_{1,1}+\sum_{e\in\pi(j_1,j_2)}(d(B))_e$,
where $\pi(j_1,j_2)$ is a path using at most $(p+q)$ edges.
Hence
\[
|B_{j_1,j_2}-B_{1,1}|
\le
\sum_{e\in\pi(j_1,j_2)}|(d(B))_e|
\le
\|d(B)\|_1
=
\|B\|_{\tv}
\le 1.
\]
Thus $|B_{j_1,j_2}|\le |B_{1,1}|+1$ for all pixels.

Now the TV norm is invariant to adding constants: $\|B+c\mathbf 1\|_{\tv}=\|B\|_{\tv}$.
Therefore, among all feasible $B$, one may w.l.o.g. shift by a constant so that
$B_{1,1}=0$ (this does not change $\|B\|_{\tv}$). Under this normalization,
$|B_{j_1,j_2}|\le 1$ for all pixels.
Hence,
\[
|\langle G,B\rangle|
=
\Big|\sum_{j_1,j_2}G_{j_1,j_2}B_{j_1,j_2}\Big|
\le
\sum_{j_1,j_2}|G_{j_1,j_2}||B_{j_1,j_2}|
\le
pq\|G\|_\infty.
\]
Taking the supremum over $\|B\|_{\tv}\le 1$ yields (i).
\end{proof}

\begin{lemma}
\label{lem:score-demo}
Under Assumption~\ref{ass:noise-design-demo}, for any $\delta\in(0,1)$ there exists
a universal constant $c>0$ such that with probability at least $1-\delta$, the
following hold simultaneously.
\begin{enumerate}
\item[(i)] Define, for each $r\in[R]$, $H_r^\ast
:=
\frac1{nT}\sum_{t=1}^T\sum_{i=1}^n \varepsilon_{ti}w_{tr}^\ast X_{ti}
\in \RR^{p\times q}$.
Then
\begin{equation}
\label{eq:HB-bound-demo}
\max_{r\in[R]}\|H_r^\ast\|_\infty
\le
c M_W M_X\sigma\sqrt{\frac{\log(2Rpq/\delta)}{nT}},
\quad
\max_{r\in[R]}\|H_r^\ast\|_{\tv}^\circ
\le
c pq M_W M_X\sigma\sqrt{\frac{\log(2Rpq/\delta)}{nT}}.
\end{equation}

\item[(ii)] Define, for each $(t,r)\in[T]\times[R]$,
$h_{tr}^\ast:=\frac1{nT}\sum_{i=1}^n \varepsilon_{ti}\langle X_{ti},B_r^\ast\rangle$.
Then
\begin{equation}
\label{eq:h-bound-demo}
\max_{t\in[T],r\in[R]}|h_{tr}^\ast|
\le
c M_B M_X\sigma\sqrt{\frac{\log(2TR/\delta)}{nT^2}}
=
c\frac{M_B M_X\sigma}{T}\sqrt{\frac{\log(2TR/\delta)}{n}}.
\end{equation}
\end{enumerate}
\end{lemma}

\begin{proof}[Proof of Lemma~\ref{lem:score-demo}]
We prove (i) and (ii) on a single event of probability at least $1-\delta$ by
combining entrywise sub-Gaussian tail bounds with a union bound.

\textit{Step 1: Proof of (i).}
Fix $r\in[R]$ and a pixel $(j_1,j_2)\in[p]\times[q]$.
Write $(H_r^\ast)_{j_1,j_2}=\sum_{t=1}^T\sum_{i=1}^n Z_{ti}$ where
$Z_{ti}:=\frac{1}{nT}\varepsilon_{ti}w_{tr}^\ast (X_{ti})_{j_1,j_2}$.
Conditional on $\{X_{ti}\}$, the random variables $\{Z_{ti}\}_{t,i}$ are
independent and mean-zero. Moreover, by Assumption~\ref{ass:noise-design-demo},
for every $u\in\RR$,
$\EE[\exp(u\varepsilon_{ti})\mid X_{ti}]\le \exp(\frac12\sigma^2u^2)$.
Therefore, still conditional on $\{X_{ti}\}$,
\[
\EE[\exp(u Z_{ti})\mid X_{ti}]
=
\EE\Big[\exp\Big(\frac{u}{nT}w_{tr}^\ast (X_{ti})_{j_1,j_2}\varepsilon_{ti}\Big)\Bigm| X_{ti}\Big]
\le
\exp\Big(
\frac12\sigma^2\Big(\frac{u}{nT}w_{tr}^\ast (X_{ti})_{j_1,j_2}\Big)^2
\Big).
\]
Using $|w_{tr}^\ast|\le M_W$ and $|(X_{ti})_{j_1,j_2}|\le\|X_{ti}\|_F\le M_X$,
we obtain
\[
\EE[\exp(u Z_{ti})\mid X_{ti}]
\le
\exp\Big(
\frac{u^2}{2}\frac{\sigma^2 M_W^2 M_X^2}{n^2T^2}
\Big).
\]
By independence across $(t,i)$ and iterating conditional expectations,
$(H_r^\ast)_{j_1,j_2}=\sum_{t,i}Z_{ti}$ is conditionally sub-Gaussian with
variance proxy at most $\frac{\sigma^2 M_W^2 M_X^2}{nT}$.
Consequently, there is a universal constant $c_0>0$ such that for all $u>0$,
\[
\PP\Big(|(H_r^\ast)_{j_1,j_2}|\ge u\Bigm|\{X_{ti}\}\Big)
\le
2\exp\Big(-\frac{c_0 nT u^2}{\sigma^2M_W^2M_X^2}\Big).
\]
Unconditioning preserves the same bound. Now apply a union bound over all
$r\in[R]$ and all $pq$ pixels:
\[
\PP\Big(\max_{r\in[R]}\|H_r^\ast\|_\infty\ge u\Big)
\le
2Rpq\exp\Big(-\frac{c_0 nT u^2}{\sigma^2M_W^2M_X^2}\Big).
\]
Set $u=c M_WM_X\sigma\sqrt{\frac{\log(2Rpq/\delta)}{nT}}$ with $c>0$ large enough so that the right-hand side is at most $\delta/2$.
This yields the stated $\|\cdot\|_\infty$ bound in \eqref{eq:HB-bound-demo}.
Finally, by Lemma~\ref{lem:tv-dual-decomp-demo}(i), for each $r$,
$\|H_r^\ast\|_{\tv}^\circ\le pq\|H_r^\ast\|_\infty$, hence
$\max_{r\in[R]}\|H_r^\ast\|_{\tv}^\circ\le pq\max_{r\in[R]}\|H_r^\ast\|_\infty$,
and we conclude the second bound in \eqref{eq:HB-bound-demo}.

\textit{Step 2: Proof of (ii).}
Fix $(t,r)\in[T]\times[R]$ and write $h_{tr}^\ast=\sum_{i=1}^n U_i$ where
$U_i:=\frac{1}{nT}\varepsilon_{ti}\langle X_{ti},B_r^\ast\rangle$.
Conditional on $\{X_{ti}\}$, the random variables $\{U_i\}_{i=1}^n$ are
independent and mean-zero. By the same argument as in part (i),
\[
\EE[\exp(uU_i)\mid X_{ti}]
\le
\exp\Big(
\frac12\sigma^2\Big(\frac{u}{nT}\langle X_{ti},B_r^\ast\rangle\Big)^2
\Big).
\]
Using Cauchy--Schwarz and $|\langle X_{ti},B_r^\ast\rangle|\le\|X_{ti}\|_F\|B_r^\ast\|_F\le M_X M_B$,
we get
\[
\EE[\exp(uU_i)\mid X_{ti}]
\le
\exp\Big(
\frac{u^2}{2}\frac{\sigma^2M_X^2M_B^2}{n^2T^2}
\Big).
\]
Thus $h_{tr}^\ast=\sum_{i=1}^nU_i$ is conditionally sub-Gaussian with variance
proxy at most $\frac{\sigma^2M_X^2M_B^2}{nT^2}$.
Therefore, for a universal constant $c_0>0$ and all $u>0$,
\[
\PP\Big(|h_{tr}^\ast|\ge u\Bigm|\{X_{ti}\}\Big)
\le
2\exp\Big(-\frac{c_0 nT^2 u^2}{\sigma^2M_X^2M_B^2}\Big).
\]
Unconditioning again preserves the inequality. Apply a union bound over all
$TR$ pairs $(t,r)$:
\[
\PP\Big(\max_{t\in[T],r\in[R]}|h_{tr}^\ast|\ge u\Big)
\le
2TR\exp\Big(-\frac{c_0 nT^2 u^2}{\sigma^2M_X^2M_B^2}\Big).
\]
Set $u=c M_BM_X\sigma\sqrt{\frac{\log(2TR/\delta)}{nT^2}}
=c\frac{M_BM_X\sigma}{T}\sqrt{\frac{\log(2TR/\delta)}{n}}$
with $c>0$ large enough so that the right-hand side is at most $\delta/2$.
This proves \eqref{eq:h-bound-demo}.

\medskip
Finally, intersecting the high-probability events from parts (i) and (ii)
gives a single event of probability at least $1-\delta$ on which both bounds
hold simultaneously.
\end{proof}

\begin{lemma}
\label{lem:int-mismatch-demo}
Define the average coefficient error
$A:=\frac1T\sum_{t=1}^T\|\widehat C_t-C_t^\ast\|_F^2$.
Let $\Delta B_r:=\widehat B_r-B_r^\ast$ and $\Delta w_{tr}:=\widehat w_{tr}-w_{tr}^\ast$.
Define the three blocks
$D^{(B)}_t:=\sum_{r=1}^R \widehat w_{tr}\Delta B_r$,
$D^{(W)}_t:=\sum_{r=1}^R \Delta w_{tr} B_r^\ast$,
$D^{(int)}_t:=\sum_{r=1}^R \Delta w_{tr}\Delta B_r$.
Let
\[
A_B:=\frac1T\sum_{t=1}^T\|D^{(B)}_t\|_F^2,
\quad
A_W:=\frac1T\sum_{t=1}^T\|D^{(W)}_t\|_F^2,
\quad
A_{int}:=\frac1T\sum_{t=1}^T\|D^{(int)}_t\|_F^2.
\]
Also define $\widetilde D^{(W)}_t:=\sum_{r=1}^R \Delta w_{tr}\widehat B_r$ and
$\widetilde A_W:=\frac1T\sum_{t=1}^T\|\widetilde D^{(W)}_t\|_F^2$.
Then the following deterministic relations hold:
\begin{enumerate}
\item[(i)] $A\le 3(A_B+A_W+A_{int})$.
\item[(ii)] $A_W\le 2\widetilde A_W+2A_{int}$,
$\frac1T\sum_{t=1}^T\big\|\sum_{r=1}^R w_{tr}^\ast\Delta B_r\big\|_F^2\le 2A_B+2A_{int}$.
\item[(iii)] $A_{int}\le 4M_B^2R\frac1T\sum_{t=1}^T\sum_{r=1}^R(\Delta w_{tr})^2
\le \frac{4M_B^2R}{m_B}\widetilde A_W$.
\end{enumerate}
\end{lemma}

\begin{proof}[Proof of Lemma~\ref{lem:int-mismatch-demo}]
\textit{Step 1: Proof of (i).}
The identity $\Delta C_t=D^{(B)}_t+D^{(W)}_t+D^{(int)}_t$ follows by expanding
$\widehat w_{tr}\widehat B_r-w_{tr}^\ast B_r^\ast$ and collecting terms.
Then use $(a+b+c)^2\le 3(a^2+b^2+c^2)$ in Frobenius norm and average over $t$.

\textit{Step 2: Proof of (ii).}
Note that
$D^{(W)}_t=\sum_r\Delta w_{tr}B_r^\ast
=\sum_r\Delta w_{tr}\widehat B_r-\sum_r\Delta w_{tr}\Delta B_r
=\widetilde D^{(W)}_t-D^{(int)}_t$.
Thus $\|D^{(W)}_t\|_F^2\le 2\|\widetilde D^{(W)}_t\|_F^2+2\|D^{(int)}_t\|_F^2$,
and averaging gives $A_W\le 2\widetilde A_W+2A_{int}$.
Similarly,
$\sum_r w_{tr}^\ast\Delta B_r
=\sum_r \widehat w_{tr}\Delta B_r-\sum_r\Delta w_{tr}\Delta B_r
=D^{(B)}_t-D^{(int)}_t$,
so the second inequality follows in the same way.

\textit{Step 3: Proof of (iii).}
First, by feasibility, $\|\Delta B_r\|_F\le \|\widehat B_r\|_F+\|B_r^\ast\|_F\le 2M_B$.
Hence for each $t$,
\[
\|D^{(int)}_t\|_F
=
\Big\|\sum_{r=1}^R \Delta w_{tr}\Delta B_r\Big\|_F
\le
\sum_{r=1}^R|\Delta w_{tr}|\|\Delta B_r\|_F
\le
2M_B\sum_{r=1}^R|\Delta w_{tr}|.
\]
By Cauchy--Schwarz, $(\sum_r|\Delta w_{tr}|)^2\le R\sum_r(\Delta w_{tr})^2$.
Thus
\[
\|D^{(int)}_t\|_F^2
\le
4M_B^2R\sum_{r=1}^R(\Delta w_{tr})^2.
\]
Average over $t$ to get the first inequality in (iii).

For the second inequality, use the Gram lower bound for $\widehat B$:
$\lambda_{\min}(G_B(\widehat B))\ge m_B$.
Let $E_W:=\frac1T\sum_{t,r}(\Delta w_{tr})^2$. Then
\[
\widetilde A_W
=
\frac1T\sum_{t=1}^T\Big\|\sum_{r=1}^R \Delta w_{tr}\widehat B_r\Big\|_F^2
\ge
m_B E_W.
\]
Hence $E_W\le \widetilde A_W/m_B$, and combining with the first inequality
gives $A_{int}\le (4M_B^2R/m_B)\widetilde A_W$.
\end{proof}

\end{document}